\documentclass[reprint,amsmath,amssymb,aps,prl]{revtex4-2}
\usepackage{graphicx}
\usepackage{color}
\usepackage{dcolumn}
\usepackage{latexsym}

\usepackage[normalem]{ulem}
\usepackage{hyperref,amssymb}
\usepackage{url}
\usepackage{color,soul}
\usepackage{graphicx}
\usepackage{verbatim}
\usepackage{multirow}
\usepackage{amsmath}
\newcommand{\beq}{\begin{eqnarray}}
\newcommand{\eeq}{\end{eqnarray}}
\usepackage{mathrsfs}
\usepackage{float,soul}
\usepackage[dvipsnames]{xcolor}
\usepackage{mathtools}
\usepackage{slashed}
\usepackage{physics}
\usepackage{graphicx}
\usepackage{epstopdf}
\usepackage{subfigure}
\usepackage[font=small]{caption}
\usepackage{hyperref}
\usepackage{bbold}
\usepackage{wasysym}
\usepackage{feynmp}
\usepackage{hyperref}
\hypersetup{colorlinks}
\usepackage{ragged2e}
\usepackage{siunitx}
\usepackage[utf8]{inputenc}

\def\red{\color{red}}

\begin{document}

\title{Wigner solid or Anderson solid—2D electrons in strong disorder}
\author{Aryaman Babbar}
\thanks{These authors contributed equally to this work}
\author{Zi-Jian Li}
\thanks{These authors contributed equally to this work}
\author{Sankar Das Sarma}
\affiliation{Condensed Matter Theory Center and Joint Quantum Institute, Department of Physics, University of Maryland, College Park, Maryland 20742, USA}

\begin{abstract}
{Critically analyzing recent STM and transport experiments \cite{ge2025visualizing} on 2D electron systems in the presence of random quenched impurities, we argue that the resulting low-density putative ``solid" phase reported experimentally is better described as an Anderson solid with the carriers randomly spatially localized by impurities than as a Wigner solid where the carriers form a crystal due to an interaction-induced spontaneous breaking of the translational symmetry.  In strongly disordered systems, the resulting solid is amorphous, which is adiabatically connected to the infinite disorder Anderson fixed point rather than the zero disorder Wigner crystal fixed point.}
\end{abstract}
\maketitle

\textit{Introduction} — Wigner pointed out in 1934, in one of the very first examples of quantum many-body phenomena, that an interacting electron gas in a neutralizing positively charged jellium background, would crystallize at low density into a ``Wigner crystal" (WC)  to optimize its Coulomb energy which dominates over the quantum zero-point fermionic kinetic energy at low density~\cite{wigner1934on}.  The realization of the quantum WC in the laboratory has been a persistent theme in condensed matter physics for $>$50 years~\cite{fertig1996properties,shayegan1996case,monarkaha20122D,shayegan2022wigner,ke2025exploring}.  

Two-dimensional (2D) semiconductors have been the most studied system in the search for quantum WC since these doped 2D materials allow a continuous change of the carrier density through gating, so the density-dependent electronic properties can be studied conveniently in a single sample to see whether a WC develops at very low density~\cite{fertig1996properties,shayegan1996case,monarkaha20122D,shayegan2022wigner,ke2025exploring}. 
Most experiments are transport studies, where the system makes a sharp transition from a good metal to a good insulator, i.e., an ``MIT", at some critical density $n_{\rm MIT}$~\cite{ahn2023density,dassarma2014two}. If this measured $n_{\rm MIT}$ turns out to be close to the theoretically expected $n_{\rm WC}$ for Wigner crystallization based on numerical calculations of the ground state energetics~\cite{drummond2009phase}, typically, a WC discovery, albeit indirect through a transport MIT, is claimed. 

There are, however, problems in interpreting such an MIT as a WC formation. First, a WC is not an insulator unless it is strongly pinned by impurities, so an MIT necessarily requires disorder.  Second, and more importantly, 2D doped semiconductors inevitably have random quenched charged impurities, and lowering the carrier density $n$ enhances the impurity effect since lowering $n$ implies enhancing $n_i/n$, with $n_i$ being the fixed sample impurity density. 
In fact, the MIT happens often around $n\sim n_i$, clearly indicating the importance of disorder~\cite{ahn2023density}! 
Thus, any claim of WC in 2D doped semiconductors must unavoidably address disorder, which is as present, if not more, in most samples, as the interaction driving the crystallization.

Most importantly, the low-density 2D MIT almost always manifests itself at a theoretical density $n_{\rm IRM}$, which is close $(n_{\rm MIT}\sim n_{\rm IRM}$) to the carrier density where the Ioffe-Regel-Mott (IRM) criterion for strong localization is satisfied, suggesting that the transition is to a strongly localized phase where the electrons are spatially randomly quenched in an amorphous solid obeying the IRM criterion~\cite{ahn2023density,dassarma2014two}. 
This is more consistent with Anderson localization, as proposed in Anderson’s landmark 1958 paper~\cite{anderson1958absence}, than with Wigner crystallization~\cite{wigner1934on}.

Occasionally, the vague terminology ``Wigner solid" (WS) is used to describe such a highly disordered amorphous phase, but perhaps ``Anderson solid" (AS) is a more appropriate terminology since disorder effects are strong (with $n\sim n_i$), and any short-range order caused by interaction is only of $O(a)$ or less where $a$ is the density dependent lattice constant, $a \sim [2/(\sqrt{3} n)]^{1/2} \sim 1.08 / n^{1/2}$, of the corresponding pristine WC. 
In other words, the ``Larkin length" is very short, of order unity~\cite{larkin1970effect}. This can be seen by realizing that the electron mean free path $l$ at $n=n_{\rm IRM}$ must obey $l<1/k_F (\sim 0.6/n^{1/2}) <a (\sim 1.08 / n^{1/2})$, which implies that any short-range order in the solid phase exists over less than one lattice constant.
The only exceptions could be cases involving extremely pure samples (with long mean free paths even at low densities), where the observed $n_{\rm MIT}$ for the 2D MIT is much lower than the corresponding (theoretically predicted) critical density $n_{\rm WC}$ for Wigner crystallization~\cite{manfra2007transport,pack2024charge,huang2024electronic}.
Then, $n_{\rm MIT}<n_{\rm WC}$ would indicate a regime of electron density $n_{\rm WC} > n> n_{\rm MIT}$, where the system is a WS, but not yet an AS. 
However, transport manifests no features at $n_{\rm WC} (\gg n_{\rm MIT} \sim n_{\rm IRM})$ in these pristine samples, thus invalidating the rationale behind using $n_{\rm MIT}$ as a transition to the WC. 
Claiming WC (or WS) simply because of the quantitative coincidence of $n_{\rm MIT} \sim n_{\rm WC}$ is misleading since the basic MIT phenomenon is the same, independent of the relative magnitudes of the experimental $n_{\rm MIT}$ and the theoretical $n_{\rm WC}$. 
The physics is always that of a metal crossing over to an insulator at $n \sim n_{\rm MIT}$ mostly because of disorder. 
Speculating that in one case (when $n_{\rm MIT}<n_{\rm WC}$) the transition is purely disorder-driven (and hence the low-density phase is AS) and in the other case (when $n_{\rm MIT}\sim n_{\rm WC}$), the low-density phase is a pinned WC (or WS) is simply a semantic exercise emphasizing a distinction without any difference.  
Thus, transport by itself cannot establish a WC, and most transport results are in fact consistent with the MIT being a transition to an AS phase as defined by the well-established IRM criterion~\cite{ioffe1960non,Mott1990MIT}.

The essence of a WC is the existence of a characteristic coherent length scale $(\sim a)$ in the solid phase, because of spontaneous symmetry breaking, implying a lattice formation, which is impossible to observe in a bulk transport measurement. It is therefore significant that several recent experiments~\cite{tsui2024direct,xiang2025imaging,ge2025visualizing} have used STM on 2D samples with variable density to directly image the carriers in the low-density solid phase as well as the transition between putative solid and liquid phases at some critical density $n\sim n_c$. (In addition, the detailed experimental transport data clearly showing an MIT is also available~\cite{ge2025visualizing,WangKimPrivateComm} for a sample very similar to the one studied in one of the STM experiments, enabling a direct comparison between transport and STM~\cite{ge2025visualizing}. )
The STM experimental critical density $n_c$ is strongly dependent on the impurity content, which is also imaged in situ in these STM studies. Curiously, although $n_c$ is typically higher than the theoretically predicted $n_{\rm WC}$ for 2D Wigner crystallization, these experiments still claim the formation of a WS.
This is attributed to a disorder-driven stabilization of the Wigner phase, whereas, as argued by us above, the $n_c>n_{\rm WC}$ phase is more likely an amorphous AS phase dominated by disorder, with $n_c \sim n_{\rm IRM}>n_{\rm WC}$\cite{vu2022thermal}, rather than a WS phase as claimed in the experiments. 

To emphasize the singularly strong role of disorder in the physics of these STM studies of our focus~\cite{xiang2025imaging,ge2025visualizing}, we mention that the solid phase (AS or WS) seen in these disordered-dominated STM experiments sometimes persists down to $r_s \sim 15$. 
Here $r_s$ is the standard dimensionless density-dependent interaction strength (the so-called Wigner-Seitz radius), defined as the inter-particle separation in units of the effective Bohr radius: $r_s = [(\pi n) ^{-1/2}]/[\kappa \hbar^2/me^2]$, where $\kappa$ and $m$ are respectively the effective background dielectric constant and effective mass.
The best current theoretical estimate for WC is: $r_s\gtrsim 38$~\cite{drummond2009phase}, but the experimentally claimed WS somehow remains stable to densities as high as $r_s \sim 15$, therefore the experimental $n_c$ is an order of magnitude larger than $n_{\rm WC}$! On the other hand, as we discuss below, $n_c \sim n_{\rm IRM}$, in these experiments, and therefore, a more natural interpretation is that the solid phase is a random AS which could be stable to very high densities in very large disorder.  It is an amorphous random solid with very short spatial ordering, in contrast to Wigner’s prediction of a periodic electron crystal. In the rest of this work, we focus on the two zero-field STM measurements on WS~\cite{xiang2025imaging,ge2025visualizing}. 

\begin{figure}[t]
    \centering
    \includegraphics[width=\linewidth]{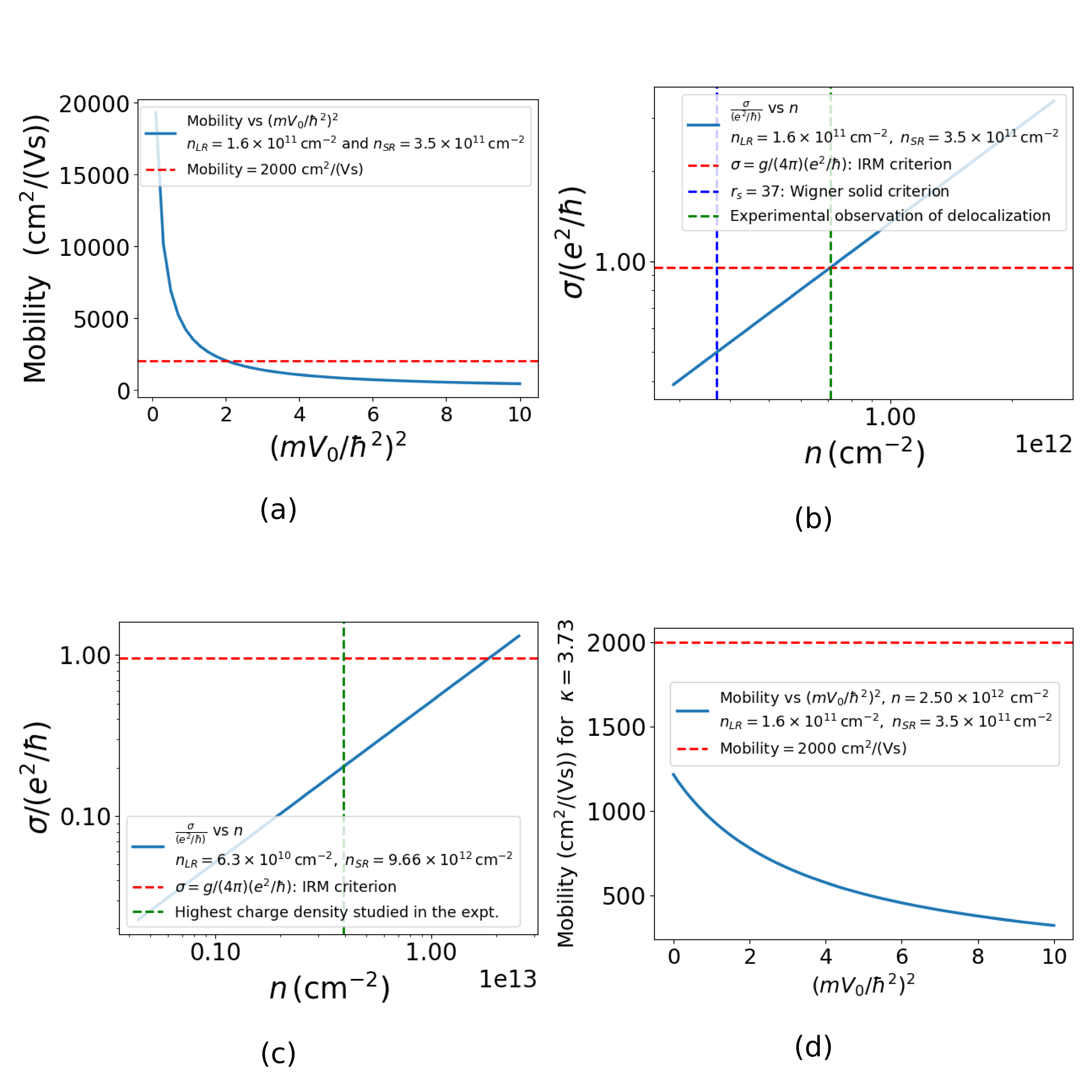}

    \caption{\justifying{ Estimation of disorder strength from experimental parameters and conductivity as a function of electron density.
    \textbf{(a)} Mobility as a function of the dimensionless parameter
    $\left(mV_0/\hbar^2\right)^2$ (a measure of the short-ranged disorder strength) given the reported values of $n_{\rm SR}$ and $n_{\rm LR}$ (blue curve). The red dotted line marks the value of the reported mobility -- 2000 $\text{cm}^2 \text{ V}^{-1} \text{ s}^{-1}$~\cite{WangPrivateComm}. The intersection is our estimate of $\left(mV_0/\hbar^2\right)^2$.
    \textbf{(b)} Conductivity for the low disorder
    density case using the values of $n_{\rm SR}$, $n_{\rm LR}$, and
    $\left(mV_0/\hbar^2\right)^2$. The blue dotted line denotes $r_S = 37$, and the green dotted line denotes the highest charge density at which localization is observed experimentally.
    \textbf{(c)} Same as \textbf{(b)} for the high disorder density case. The green dotted line denotes the highest charge density reported in the experiment.
    In both panels \textbf{(b)} and \textbf{(c)}, the red dotted line denotes the IRM criterion, and the blue solid line corresponds to the conductivity.
    \textbf{(d)} Failure of the assumption $g_v = 1$: Mobility as a function of $\left(mV_0/\hbar^2\right)^2$ with $n = 2.5 \times 10^{12} \text{ cm}^{-2}$ (blue curve), does not intersect $\mu = 2000 \text{ cm}^2 \text{ V}^{-1} \text{ s}^{-1}$ for $\left(mV_0/\hbar^2\right)^2 > 0$.
    }
    }
    \label{fig:transport_figure_1}
\end{figure}

\begin{figure}[t]
    \centering
    \includegraphics[width=\linewidth]{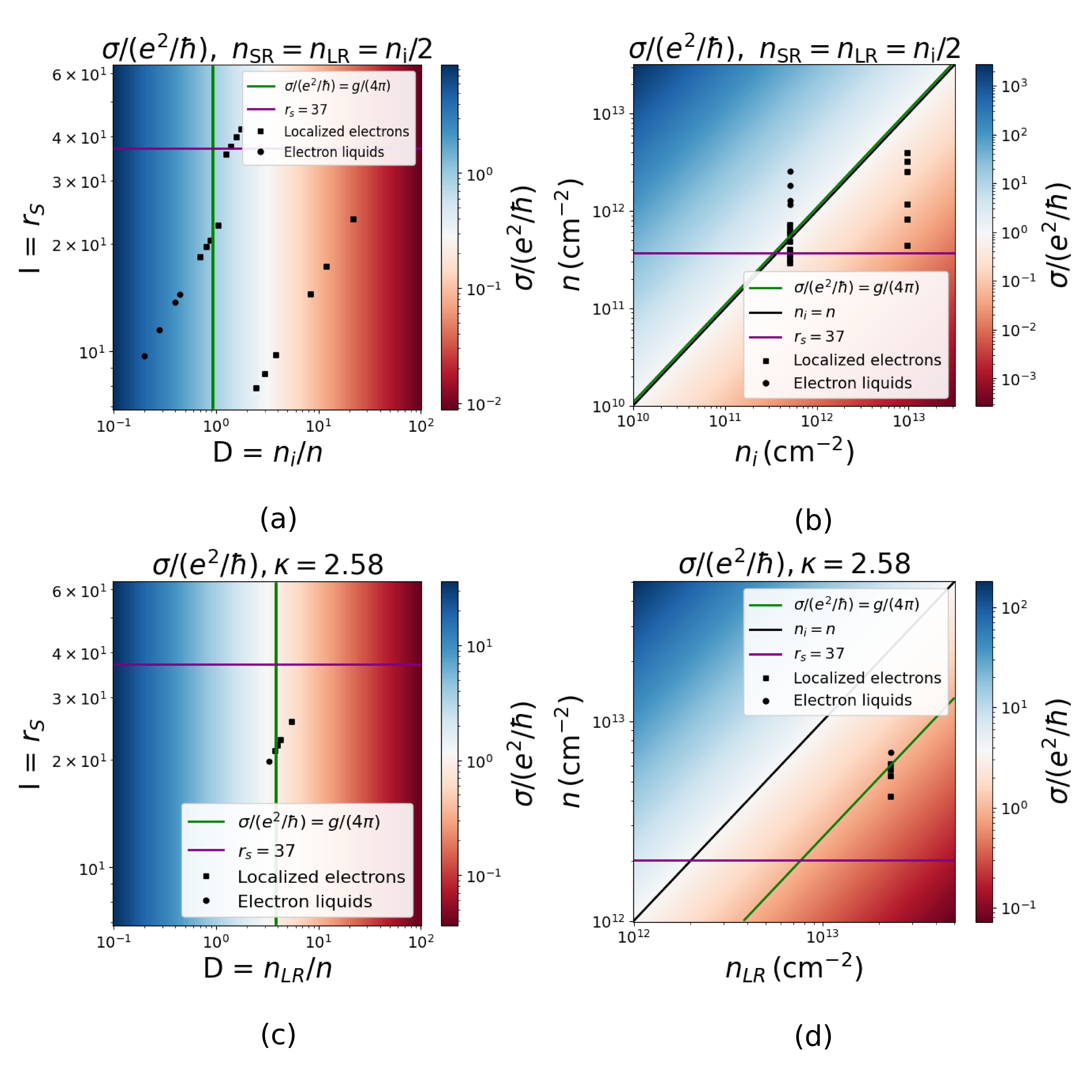}

    \caption{\justifying{ Transport phase diagrams as functions of experimental disorder parameters.
    \textbf{(a)} $\sigma/\left(e^2/\hbar\right)$ as a function of $r_s$ and $n_i/n$, where $\sigma$ is the conductivity,
    $n_i = n_{\rm SR} + n_{\rm LR}$, and we assume $n_{\rm SR} = n_{\rm LR}$.
    Experimentally reported $(r_s, n_i/n)$ pairs are shown as squares (electron solids) or circles (electron liquids)~\cite{ge2025visualizing, WangPrivateComm}.
    \textbf{(b)} $\sigma/\left(e^2/\hbar\right)$ as a function of
    $n_i = n_{\rm SR} + n_{\rm LR}$ and $n$ on a logarithmic scale, again assuming
    $n_{\rm SR} = n_{\rm LR}$.
    Experimentally reported $(n_i, n)$ pairs are shown as squares (electron solids) or circles (electron liquids)~\cite{ge2025visualizing, WangPrivateComm}.
    \textbf{(c)} Same as \textbf{(a)} for an earlier experiment~\cite{xiang2025imaging} on the same materials with holes
    as charge carriers, assuming a mobility of
 $240\,\mathrm{cm}^2\,\mathrm{V}^{-1}\,\mathrm{s}^{-1}$, where only $n_{\rm LR}$ contributes to $n_i$.
    \textbf{(d)} Same as \textbf{(b)} for the earlier experiment~\cite{xiang2025imaging}. In all figures, the green line denotes the IRM criterion, while the purple line corresponds to $r_S = 37$. For \textbf{(b)} and \textbf{(d)}, the black line represents $n_i = n$.
    }
    }
    \label{fig:transport_figure_2}
\end{figure}

\textit{Transport}—We start by showing our transport results in Figs.~\ref{fig:transport_figure_1} and~\ref{fig:transport_figure_2} for the two recent STM experiments, on 2D holes and electrons. 
(Details of the transport theory are given in Supplementary Materials~\cite{SM}, but the theory is standard and has been extensively covered in the literature (see, e.g., Ref.~\cite{huang2024electronic}.)) 

We provide a transport ``phase diagram" based on the application of the IRM criterion on the calculated density-dependent conductivity, which we obtain from the known sample mobility~\cite{ge2025visualizing,WangKimPrivateComm, WangPrivateComm} and/or the directly STM-imaged carrier density and impurity density.
Our focus and most of our transport results are on the very recent 2D electron STM measurements~\cite{ge2025visualizing} (along with the detailed transport data~\cite{ge2025visualizing,WangKimPrivateComm} on a very similar sample as the one used in the STM experiment).
But our conclusions apply equally well, if not more, to the 2D hole sample, which, being more strongly disordered, is deeper into the AS phase~\cite{xiang2025imaging}. 

In Fig.~\ref{fig:transport_figure_1}, we establish the quantitative disorder strength directly using the experimental parameters, where we further use this disorder strength to calculate the conductivity.  
In Fig.~\ref{fig:transport_figure_1}(a), we use the experimentally provided (through direct STM imaging) impurity density parameters for long-range (LR) (Coulomb), $n_{\rm LR}$,  and short-range (SR) (defect), $n_{\rm SR}$, parameters to fix the SR scattering strength from the measured mobility ($\sim 2000\, {\rm cm^2/(V\cdot s)}$), which is experimentally density-independent \cite{WangKimPrivateComm}. 
Although the SR-scattering-induced mobility is by definition density independent, the LR-scattering typically has a density dependence~\cite{dassarma2014two}, except for the very strong screening in the experimental system with a large effective mass ($m=0.54 m_e$) and large degeneracy ($g=g_sg_v=12$) arising from a valley degeneracy ($g_v$) of $6$ leads to a large screening constant $q_{\rm TF}=gme^2/(\kappa \hbar^2)$ which satisfies the strong screening condition $q_{\rm TF} \gg 2k_{\rm F}$ for all electron densities with the Fermi wavenumber $k_{\rm F} =(\pi n/3) ^{1/2}$ being suppressed by the large degeneracy. 
In the experimental density range, $q_{\rm TF}/k_{\rm F}>100$, and hence the screened Coulomb disorder is simply given by $u=2\pi e^2/q_{\rm TF}$ which is momentum independent, and is thus also short-ranged.  The theoretical mobility, $\mu$, which is density-independent because all disorder is effectively short-ranged, then depends on only one unknown parameter, namely the effective strength of the SR scattering:
\begin{equation}
    \mu = \frac{e \tau_t}{m}=\frac{e}{\hbar}\frac{1}{n_{\rm LR}\left(\frac{2\pi}{g}\right)^2 + n_{\rm SR}\left(\frac{mV_0}{\hbar^2}\right)^2}
\end{equation}
where $V_0$ is the unknown SR disorder strength. 
Using $\mu=2000 {\rm \, cm^2/(V\cdot s)}$, we obtain $V_0$, and hence the full disorder strength is known. 
We then calculate the conductivity, as shown in Figs. \ref{fig:transport_figure_1}(b) (low disorder density so-called LDD regime in the experiment) and \ref{fig:transport_figure_1}(c) (high disorder density so-called HDD regime), and show the characteristic calculated conductivity for strong localization (red), defined by the IRM criterion $k_{\rm F} l =1$, as well as the $n_{\rm WC}$ location (blue). 
The experimental solid-liquid transition (green) in the LDD regime happens precisely at our calculated IRM density $n_{\rm IRM}$, far above the WC critical density $n_{\rm WC}$, directly supporting the quenched random Anderson localized character of the solid phase. 
In the HDD regime, the IRM regime (below the red line) spans the whole electron density range studied by STM, and everything is in the AS phase, in agreement with the STM data.  In Fig.~\ref{fig:transport_figure_1}(d), we show that an assumption of $g_v=1$ manifestly fails to account for a sample mobility of $2000 {\rm\, cm^2/(V\cdot s)}$ using the experimental values of $n_{\rm SR}$ and $n_{\rm LR}$ and $n$, since $V_0^2$ must become ``negative"(!), i.e., disorder enhancing the conductivity, to explain the measured mobility.  A $g_v=1$ implies very strong LR disorder scattering, which is incompatible with the data.

In Fig.~\ref{fig:transport_figure_2}, we provide a summary of our extensive transport results by depicting transport phase diagrams (red: AS; blue: Metallic Fermi liquid (FL)) as functions of experimental disorder parameters, depicting the experimentally reported solid and liquid results. 
In Fig.~\ref{fig:transport_figure_2}(c), we show the same results for a hole sample with very low mobility studied by STM in an earlier experiment~\cite{xiang2025imaging}.  
In all the results, the experimental ``solid" data points are in the red AS phase, and all the ``liquid" data points are in the blue metallic FL phase, reinforcing our claim that the experimentally observed ``solid" is mostly an amorphous AS phase with very short-range spatial correlations.

A crucial point to note is that the transport MIT, commonly associated with the WC transition, often happens at $n\sim n_i$, where $n_i$ is the effective 2D random impurity density~\cite{ge2025visualizing}.
This is also consistent with our Fig.~\ref{fig:transport_figure_2}(b), where the $n_i=n$ line basically coincides with the IRM line. 
If we use $n=n_i=n_c$ defining the transition to the solid phase to be at $r_s=r_i$ (with $r_i$ defined through $n_i$), rather than the pristine WC condition $r_s=r_{\rm WC} \sim 38$, we find that all the solid phase data points of Ref.~\cite{ge2025visualizing} are consistent with having the disorder-dependent $r_i$ being the approximate transition point! 
This is compelling support for our central claim that disorder plays a dominant role in the transition, leading to a random AS rather than a periodic WS.

\begin{figure}
    \centering
    \includegraphics[width=\linewidth]{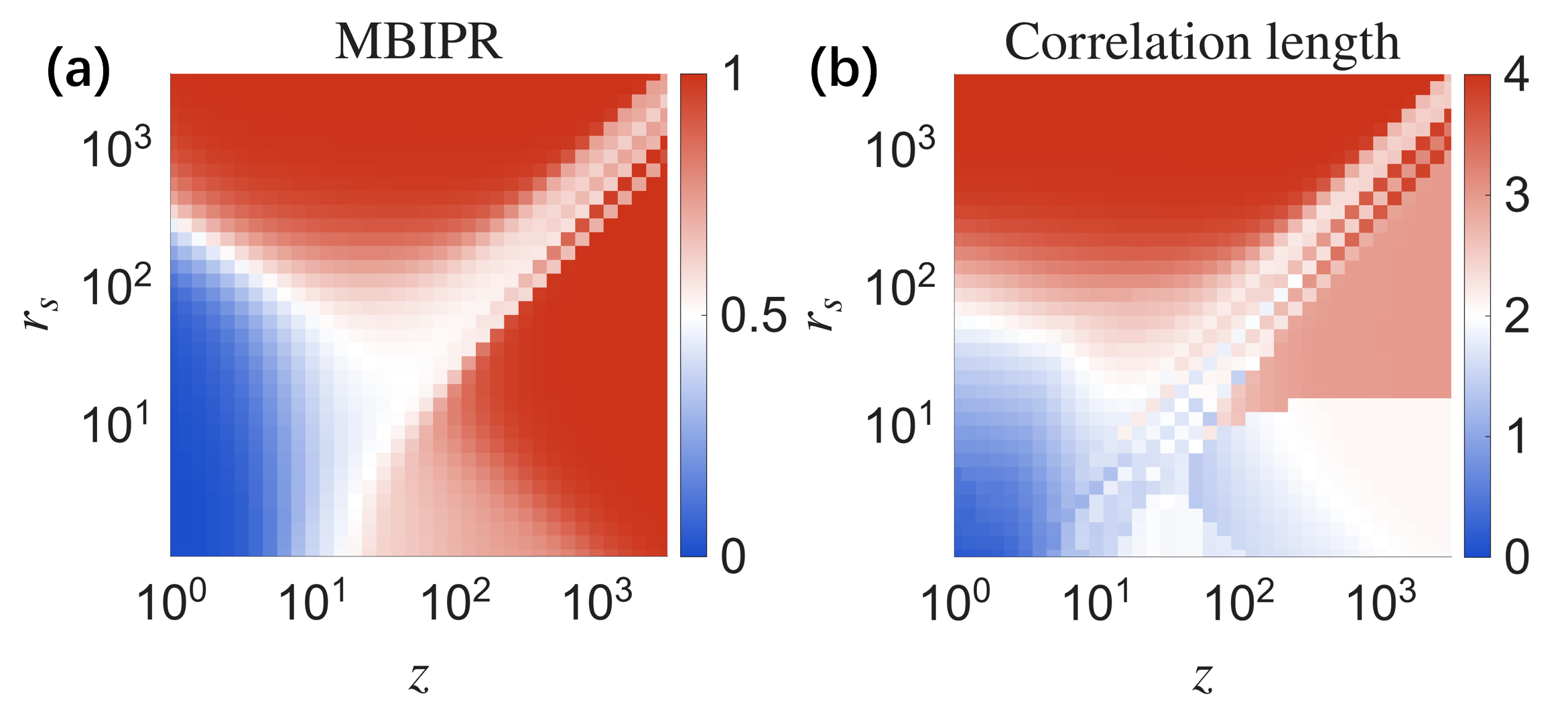}
    \caption{\justifying{Long-range disorder phase diagram characterized by \textbf{(a)} many-body inverse participation ratio and \textbf{(b)} effective correlation length. $z$ represents disorder strength and $r_s$ is the Wigner-Seitz radius.}}
    \label{fig:LR}
\end{figure}

\begin{figure}
    \centering
    \includegraphics[width=\linewidth]{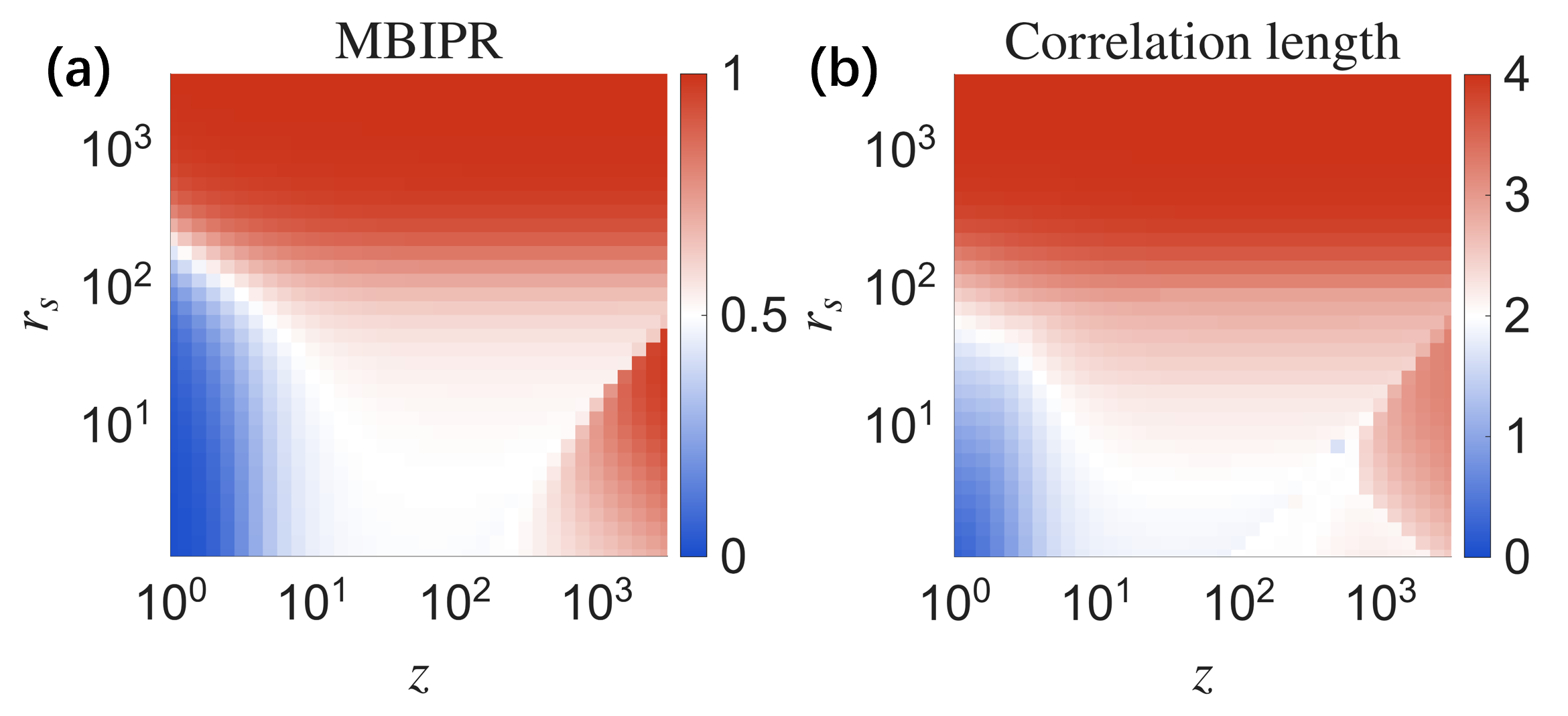}
    \caption{\justifying{Short-range disorder phase diagram characterized by \textbf{(a)} many-body inverse participation ratio and \textbf{(b)} effective correlation length. $z$ represents disorder strength and $r_s$ is the Wigner-Seitz radius.}}
    \label{fig:SR}
\end{figure}
\textit{Phase diagram} — We use the exact diagonalization (ED) technique to directly calculate the many-body phase diagram characterized by electron localization and spatial correlations for small interacting disordered electron systems.
These results are shown in Figs.~\ref{fig:LR} (LR disorder) and~\ref{fig:SR} (SR disorder).  While our transport results use realistic parameters and are directly comparable to the experimental data, the ED results are for a model Hamiltonian with variable disorder ($z$) and interaction  ($r_s$)  strengths (see Supplementary Materials for the details of the model~\cite{SM}).  

The phase diagram, both for long-range and short-range disorder (the electron-electron interaction is always long-ranged in the theory), manifests three distinct phases as characterized by the many-body inverse participation ratio (MBIPR) (see Figs.~\ref{fig:LR}(a) and~\ref{fig:SR}(a)), which provides a direct measure of the size of the electron wave packet, in other words, the magnitude of localization of the electrons. This quantity has been widely used to investigate many-body localization (e.g., Refs.~\cite{bera2015manybody,vu2022fermionic}).
The blue region for weak interaction and weak disorder is the metallic FL  phase, where the electrons are simply spread out over the system. 
The red region at the top for strong interaction and (relatively) weak disorder, which is separated from the blue region by a sharp transition (qualitatively corresponding to $n_{\rm WC}$) at a critical $r_s$, is the WC region where the electrons are spatially localized by interactions. 
But, the second red region in the lower right-hand corner for strong disorder and (relatively) weak interaction, separated by a sharp transition (qualitatively corresponding to $n_{\rm IRM}$), is the Anderson localized region where the electrons are randomly spatially localized by disorder. 

MBIPR by itself cannot distinguish the Wigner and Anderson phases (hence, both are red in the figures) since both have localized wavefunctions, distinct from the extended states in the blue FL phase. 
But our calculated effective spatial correlation length, shown in Figs.~\ref{fig:LR}(b) and~\ref{fig:SR}(b), clearly bring out the distinction between the disorder-localized AS, with relatively short correlations, and the WS, with longer correlations—of course,  the FL phase has the shortest correlations, but the AS is often closer to the liquid phase than to the Wigner phase in terms of spatial correlations, particularly for stronger disorder. 
This reinforces the fact that the AS phase is an amorphous quenched phase because of the dominance of disorder. 

We mention some salient features of our phase diagrams, which are consistent with certain reported peculiarities in the STM experiments. 
First, intermediate disorder slightly enhances the WS phase, with the effect being stronger for the LR disorder.  This manifests in the curvature of the top Wigner red region for intermediate disorder. 
Second, there is an intriguing re-entrance for strong disorder, where decreasing interaction from the large-$r_s$ at fixed large disorder first produces an almost-liquid phase (the ``white" stripe in between the two red regions for large $r_s$ and large z) at lower $r_s$, but then it re-enters a solid phase again as $r_s$  decreases further. 
This interesting competition/interplay between interaction/disorder is more manifest for the LR disorder, and this is also the experimental finding where STM studies indicate a re-entrance with decreasing $r_s$ from a solid phase at large $r_s$ to a solid phase at lower $r_s$, going through an intermediate liquid phase. 
The system is basically going from a WS  for $r_s>z$ to an AS  at $r_s<z$  through a liquid-like amorphous phase with little spatial correlations.

\textit{Conclusion} — We conclude by pointing out some connections between our transport  (Figs.~\ref{fig:transport_figure_1} and~\ref{fig:transport_figure_2}) and ED (Figs.~\ref{fig:LR} and~\ref{fig:SR}) phase diagrams. 
The purple WC horizontal line at $r_s \sim 38$ in Fig.~\ref{fig:transport_figure_2} is simply a horizontal line at $r_s \sim 38$ in Fig.~\ref{fig:LR} (also~\ref{fig:SR}) which can be drawn along the bottom of the red WS regions Figs.~\ref{fig:LR}/\ref{fig:SR}. 
(Note that $r_s$ in the ED uses arbitrary units~\cite{SM}.)
This is simply the putative first-order phase transition between the FL and the WC at zero disorder, which remains mostly unaffected until the disorder becomes very strong ($z>r_s$) in Fig.~\ref{fig:LR}/\ref{fig:SR}. 
Similarly, the green IRM line in Fig.~\ref{fig:transport_figure_2}, separating the metal and the insulator, can be drawn as a vertical line (at $z \sim 20$) separating the blue FL from the red AS regions of Figs.~\ref{fig:LR}/\ref{fig:SR}.  
Our transport analysis based on the actual experimental mobility and conductivity indicates that most of the STM experimental parameter space explores the amorphous AS phase, where disorder effects are strong, and our ED phase diagrams demonstrate nontrivial interplay between interaction and disorder when both are strong.

Finally, we mention that our work, coupled with these STM studies, suggest the possible existence of four distinct fixed points in disordered interacting electron systems:  low-disorder weakly-interacting Fermi liquid (the ``blue" phase in the bottom left corner of Figs.~\ref{fig:LR}/\ref{fig:SR}); low-disorder strongly-interacting Wigner phase (the ``red" phase at the top); high-disorder weakly-interacting Anderson phase (the ``red" phase at the bottom right); high-disorder strongly-interacting strong-coupling fixed point representing an electron glass with strong interaction induced short-range-order (the ``white" stripe around top right hand corner).
While the experimental data points in these zero-field STM experiments fall mostly (but not entirety) in the AS phase  (Figs.~\ref{fig:transport_figure_1} and~\ref{fig:transport_figure_2}), another impressive recent STM experiment in 2D bilayer graphene~\cite{tsui2024direct} under a strong magnetic field clearly establishes that the WC phase evolves into a strongly disordered AS phase with increasing effective disorder, in qualitative agreement with our analysis. 
The AS phase in this graphene experiment (at Landau level filling factor $<1/10$), as well as in the zero field experiments we discuss manifest STM images whose structure factor is featureless, representing a random amorphous solid. 
All of this is also in agreement with the Imry-Ma argument~\cite{imry1975random} of disorder destroying all long-range order in 2D systems with random domain formation, with a Larkin length of $O(1)$ because the disorder is as important as the interaction. 
The STM experiments~\cite{xiang2025imaging,ge2025visualizing} are exploring a highly fragmented WS landscape because of strong disorder, and the resulting solid is a random pinned glass rather than a Wigner crystal, in spite of strong interactions. This solid is more of an electron glass induced by random charged impurities with no spontaneous breaking of the translational symmetry, since the translational symmetry is broken explicitly by the random impurities.  This is the classic ``random field" problem with no Goldstone modes in the solid phase~\cite{imry1975random}. We emphasize that one direct consequence of the presence of the quenched random impurities in the system is that the ``solid" phase persists to very high temperatures, much higher than the melting temperature for the corresponding pristine WC~\cite{vu2022thermal}, because the impurities pin the fragmented solid, preventing melting. 
In fact, the WC melting temperature is estimated to be $\sim 0.5\text{K}$~\cite{vu2022thermal,huang2024electronic} for Ref.~\cite{ge2025visualizing}, whereas the experiment itself is done at $T>5\text{K}$, supporting our disorder-induced random AS interpretation.

\textit{Acknowledgements}—The authors gratefully acknowledge helpful communications with Feng Wang, Haleem Kim, and Zhehao Ge of UC, Berkeley.  This work is supported by the Laboratory for Physical Sciences  (LPS) through the Condensed Matter Theory Center  (CMTC) at Maryland. AB thanks the Joint Quantum Institute at the University of Maryland for support through a JQI graduate fellowship. 

\bibliographystyle{apsrev4-1}
\bibliography{ref}

\newpage
\onecolumngrid
\appendix

\clearpage
\renewcommand\thefigure{S\arabic{figure}}    
\setcounter{figure}{0} 
\renewcommand{\theequation}{S\arabic{equation}}
\setcounter{equation}{0}
\renewcommand{\thesubsection}{SM\arabic{subsection}}
\section{\LARGE Supplementary Information}

\subsection{\Large \uppercase\expandafter{\romannumeral1}. Transport in the Bilayer MoSe$_2$ sample}

We shall use Boltzmann transport theory at $T=0$ to compute the scattering rate of our charge carriers. Our charge carriers are electrons with charge $-e$ and effective mass $m = 0.54 \, m_e$. We assume that there are two kinds of disorder in our system. We have the in-plane charged (long-range) disorder where each charged impurity is of charge $e$, and we have in-plane short-range disorder of disorder strength $V_0$. In order to understand the effect of the long-range disorder, we shall use RPA, and we shall assume that both long- and short-range disorder are distributed randomly and independently within the plane of the material. The material for this experiment is Bilayer $\text{MoSe}_2$, which has a valley degeneracy of 6 and a spin degeneracy of 2, bringing the total degeneracy $g = 12$ \cite{ge2025visualizing}.

\section{\large \uppercase\expandafter{\romannumeral1}.1 Scattering rates} 

In this section, we shall use Boltzmann transport theory at $T=0$ to compute the transport and quantum scattering rate for the electrons scattered by charged (long-ranged) impurities and neutral impurities in the system.

\subsection{\large \uppercase\expandafter{\romannumeral1}.1.1 Charged disorder} 

The calculation of scattering rate for long-ranged disorder using RPA has been done in the literature, and we shall quote the result from \cite{Disorder_Si_Ge}. The transport scattering rate given a long-ranged disorder density of $n_{\rm{LR}}$ is 
\begin{equation}
    \frac{1}{\tau_t} = n_{\rm LR} \frac{2\pi \hbar}{m} \left(\frac{2}{g}\right)^2 \int_0^1 \text{d}x \, \frac{2s^2 x^2}{\sqrt{1-x^2} \left(x+s\right)^2},
\end{equation}
where
\begin{equation}
    s = \frac{q_{TF}}{2 k_F} = \frac{ge^2 m}{2 \kappa k_F \hbar^2}.
\end{equation}
$\kappa$ is relative permittivity, and $k_{\rm{F}}$ is the Fermi wavenumber. The highest electron density in our analysis is $<5 \times 10^{12} \, \text{cm}^{-2}$, $k_F = \sqrt{4 \pi n/g}$, where $n$ is the electron density, and the lowest $\kappa$ is 2.58. For these parameters, 
\begin{equation}
    \left(\frac{q_{TF}}{2 k_F}\right)_{\rm{experiment}} > \frac{12 \times \left(4.8 \times 10^{-10}\right)^2 \times 0.54 \times 9.11 \times 10^{-28}}{2 \times 2.58 \times \sqrt{\frac{4\pi \times 5 \times 10^{12}}{12}} \times (1.05 \times 10^{-27})^2} \simeq 100 \gg 1.
\end{equation}
This means that for all our densities and relative permittivities in the experiment, we can safely approximate
\begin{equation}
    \frac{1}{\tau_t} = \frac{\hbar n_{\rm LR}}{m} \left(\frac{2 \pi}{g}\right)^2.
\end{equation}
For the quantum scattering time, we have
\begin{equation}
    \frac{1}{\tau_q} = n_{\rm LR} \frac{2\pi \hbar}{m} \left(\frac{2}{g}\right)^2 \int_0^1 \text{d}x \, \frac{s^2}{\sqrt{1-x^2} \left(x+s\right)^2} \simeq \frac{\hbar n_{\rm LR}}{m} \left(\frac{2 \pi}{g}\right)^2.
\end{equation}
Both scattering times turn out to be the same, ignoring terms of $\mathcal{O}\left(s^{-1}\right)$.

\subsection{\large \uppercase\expandafter{\romannumeral1}.1.2 Short-ranged disorder} 

For short-ranged neutral disorder, we assume that the matrix element for scattering from one plane wave state to another is $\langle \vec{k}'|V|\vec{k}\rangle = V_0$, and then using Fermi's Golden rule, we obtain that both the transport and the quantum scattering rate, for $n_{\rm SR}$ the density of short-ranged disorder, in the case of a 2D material, are given by

\begin{equation}
    \frac{1}{\tau_{t/q}} = \frac{\hbar}{m} n_{\rm SR}\left(\frac{m V_0}{\hbar^2}\right)^2.
\end{equation}
We must note that this is valid for low enough electron density, which is assumed to hold for our system. The scattering times add according to Matthiessen's rule.

\section{\large \uppercase\expandafter{\romannumeral1}.2 Mobility} 

Mobility is given by

\begin{equation}
    \mu = \frac{e \tau_t}{m} = \frac{e}{\hbar} \frac{1}{n_{\rm LR} \left(\frac{2\pi}{g}\right)^2 + n_{\rm SR}\left(\frac{m V_0}{\hbar^2}\right)^2}.
\end{equation}
We seek to find the dimensionless parameter $\left(\frac{m V_0}{\hbar^2}\right)^2$, so we shall use the fact that the mobility for the low disorder density (LDD) case ($n_{\rm SR} = 3.5 \times 10^{11} \text{cm}^{-2}, \, n_{\rm LR} = 1.6 \times 10^{11} \text{cm}^{-2}$) is measured to be $\simeq 2000 \, \text{cm}^2 \, \text{V}^{-1} \, \text{s}^{-1}$ \cite{WangPrivateComm}. Plotting a graph of mobility against $\left(\frac{m V_0}{\hbar^2}\right)^2$, we find that the value that gives our required mobility is $\left(\frac{m V_0}{\hbar^2}\right)^2 \simeq 2$ (Figure \ref{fig:transport_figure_1}a).

We shall use this value for the rest of our work. Upon fixing the dimensionless parameter, our mobility is given by

\begin{equation}
    \mu \simeq \frac{e}{\hbar} \frac{1}{n_{\rm LR} \left(\frac{2\pi}{g}\right)^2 + 2 n_{\rm SR}}.
\end{equation}

\begin{figure}[t]
    \centering
    \includegraphics[width=0.45\columnwidth]{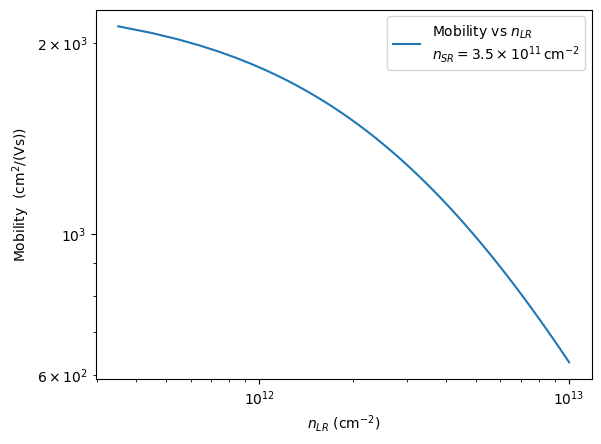}
    \hfill
    \includegraphics[width=0.45\columnwidth]{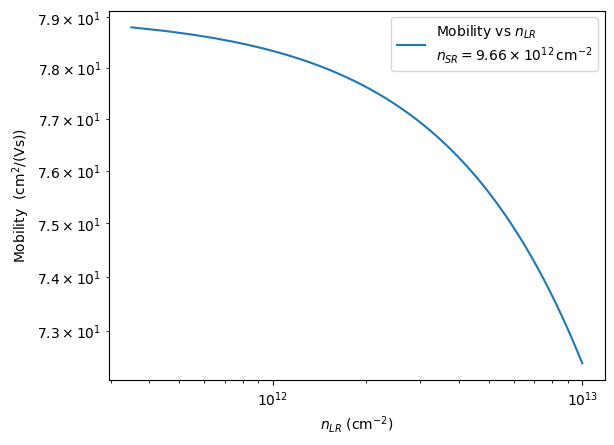}
    \caption{
    Left: mobility for $n_{\rm SR} = 3.5 \times 10^{11} \text{cm}^{-2}$ and $n_{\rm LR}$ is varied.
    Right: mobility for $n_{\rm SR} = 9.66 \times 10^{12} \text{cm}^{-2}$ and $n_{\rm LR}$ is varied.}
    \label{fig:mobvlr}
\end{figure}

We plot the mobility as a function of $n_{\rm{LR}}$ by fixing $n_{\rm{SR}}$ to its LDD and high disorder density (HDD) values (Figure \ref{fig:mobvlr}).

\begin{figure}[t]
    \centering
    \includegraphics[width=0.45\columnwidth]{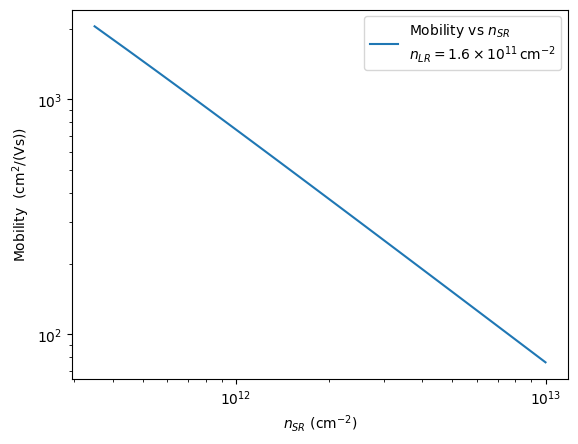}
    \hfill
    \includegraphics[width=0.45\columnwidth]{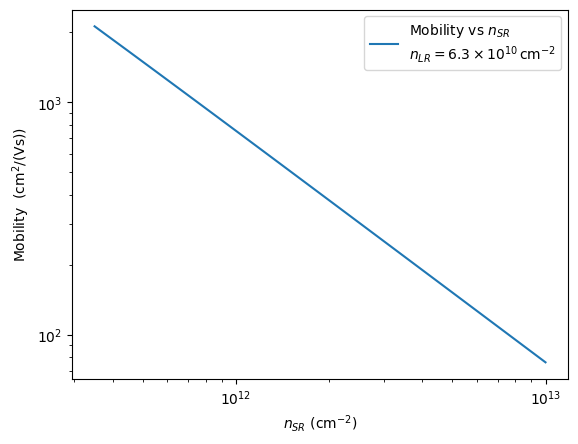}
    \caption{
    Left: mobility for $n_{\rm LR} = 1.6 \times 10^{11} \text{cm}^{-2}$ and $n_{\rm SR}$ is varied.
    Rigth: mobility for $n_{\rm LR} = 6.3 \times 10^{10} \text{cm}^{-2}$ and $n_{\rm SR}$ is varied.}
    \label{fig:mobvsr}
\end{figure}

We repeat the same process as above, only this time we fix the values of $n_{\rm{LR}}$ to its values in the LDD and HDD case to compute mobility as a function of $n_{\rm{SR}}$ (Figure \ref{fig:mobvsr}).

\section{\large \uppercase\expandafter{\romannumeral1}.3 Conductivity} 

We have that the conductivity is given by

\begin{equation}
    \sigma = \mu n e \simeq \frac{e^2}{\hbar} \frac{n}{n_{\rm LR} \left(\frac{2\pi}{g}\right)^2 + 2 n_{\rm SR}},
\end{equation}

The IRM criterion $\tau \frac{h k_F^2}{m} = 1$, directly translates to $\sigma = \frac{g}{4 \pi} \frac{e^2}{\hbar}$

We plot $\sigma$ against $n$ for the LDD and HDD case (Figures \ref{fig:transport_figure_1}b and \ref{fig:transport_figure_1}c). In the LDD case, we find that the transition happens at a higher charge density than if the transition were due to Wigner crystallization. The highest charge density reported in the experiment, which is an electron solid, is found to be in very good agreement with the value predicted by the IRM criterion. 

As for the HDD case, even the highest charge densities the experiment goes up to show localization, which is indeed seen in our plot (Figure \ref{fig:transport_figure_1}c), where this density is seen deep in the Anderson localized regime.

\section{\large \uppercase\expandafter{\romannumeral1}.4 More plots for $g=12$} 

\begin{figure}[htbp]
    \centering
    % --- First row ---
    \includegraphics[width=0.32\columnwidth]{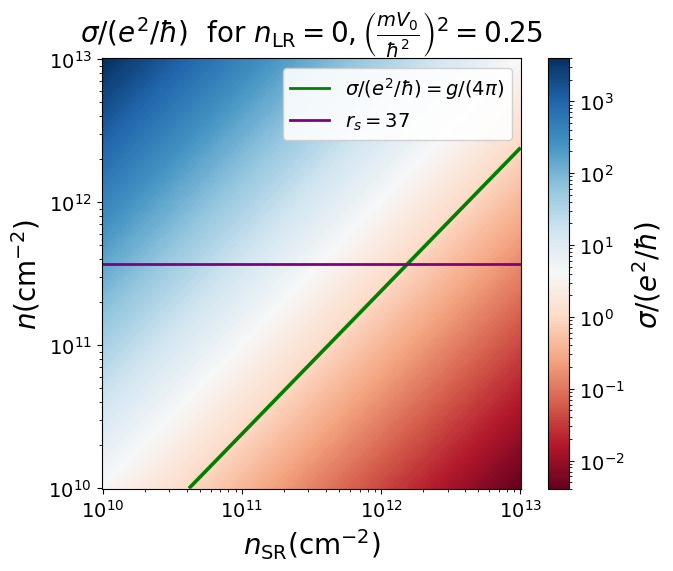}
    \includegraphics[width=0.32\columnwidth]{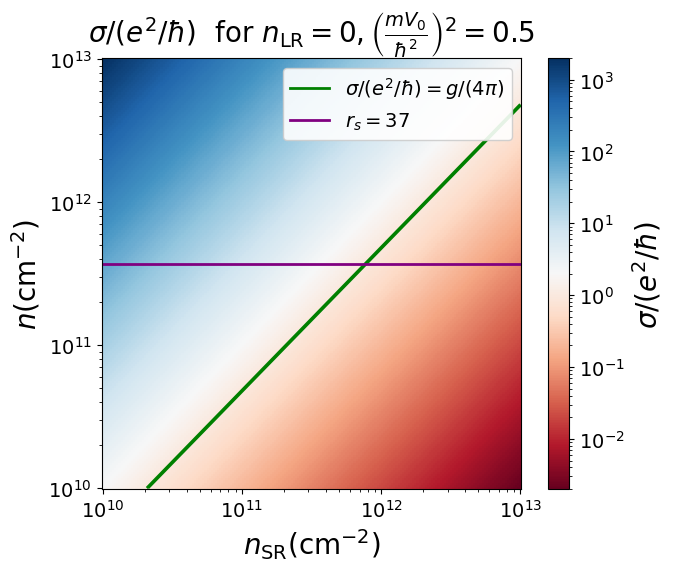}
    \includegraphics[width=0.32\columnwidth]{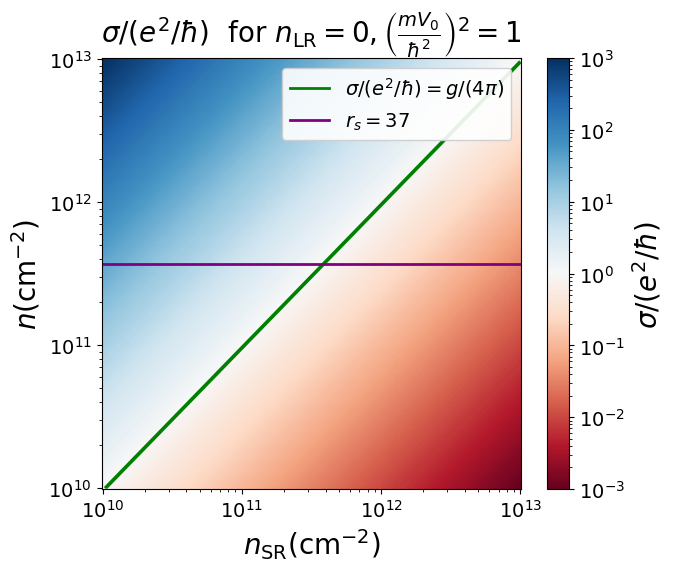}

    \vspace{2pt}

    % --- Second row ---
    \includegraphics[width=0.32\columnwidth]{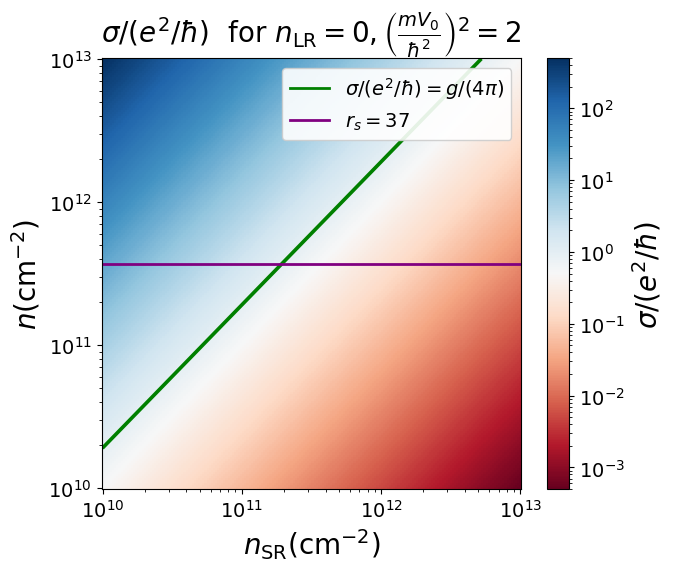}
    \includegraphics[width=0.32\columnwidth]{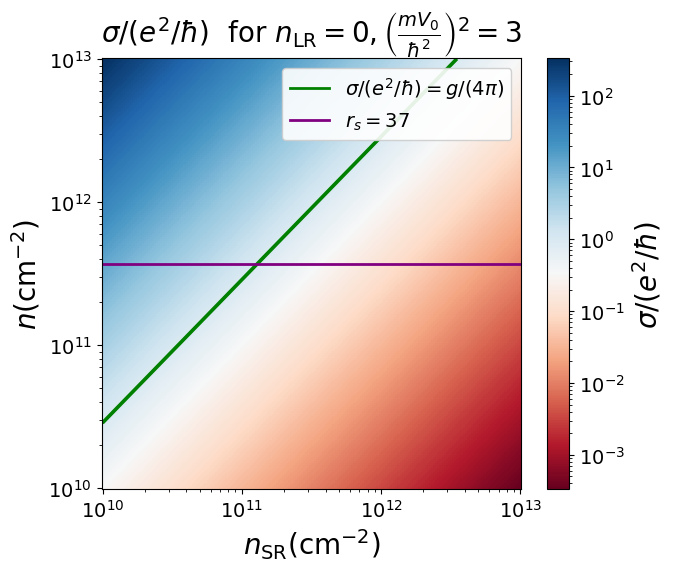}
    \includegraphics[width=0.32\columnwidth]{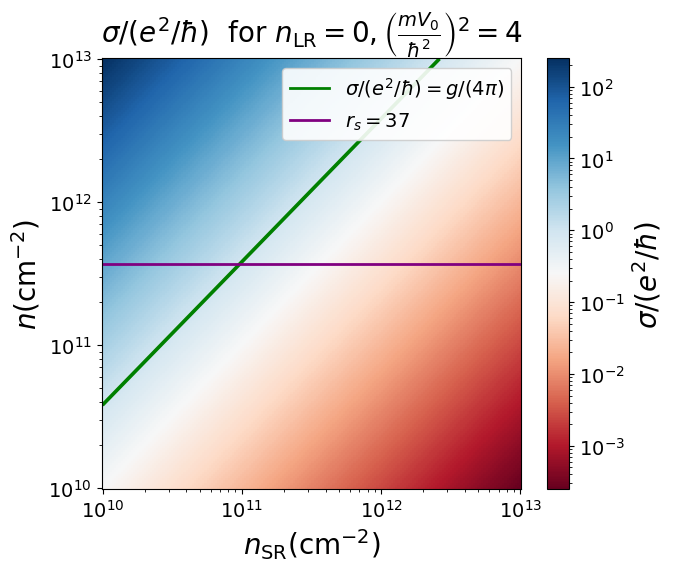}

    \caption{Conductivity maps for various fixed $\left(\frac{mV_0}{\hbar}\right)^2$, $g = 12$ and $n_{\rm LR} = 0$ for varying $n$ and $n_{\rm SR}$}
    \label{fig:nlr0parfixedvarynandnsr}
\end{figure}

\begin{figure}[htbp]
    \centering
    % --- First row ---
    \includegraphics[width=0.32\columnwidth]{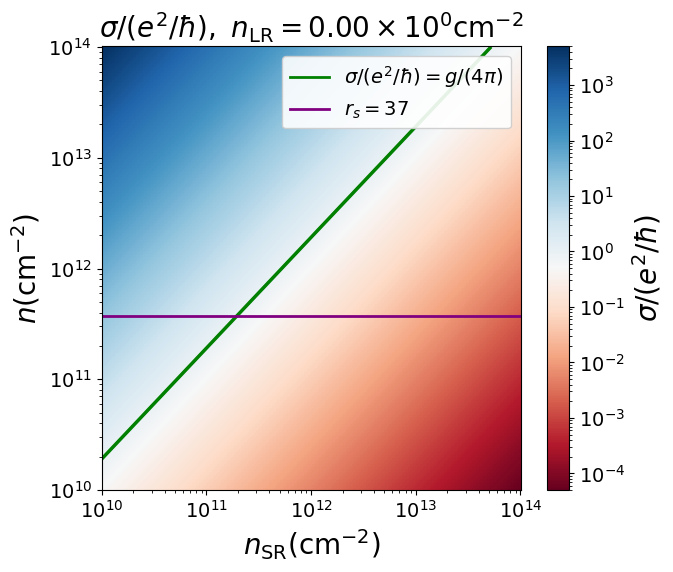}
    \includegraphics[width=0.32\columnwidth]{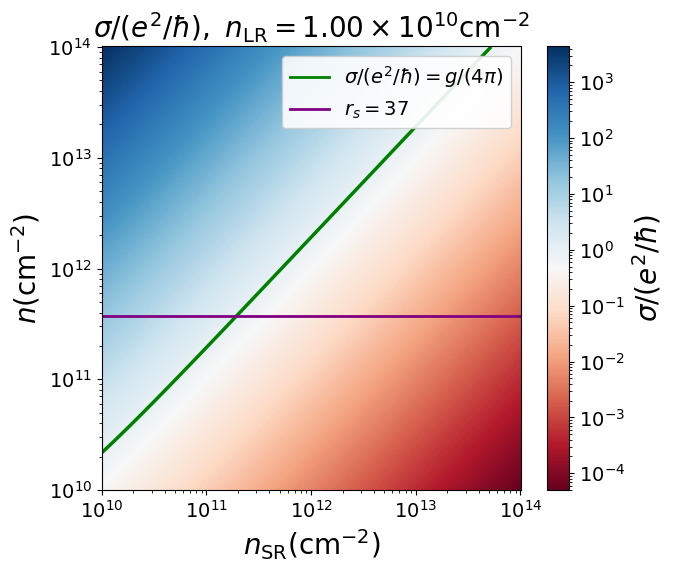}
    \includegraphics[width=0.32\columnwidth]{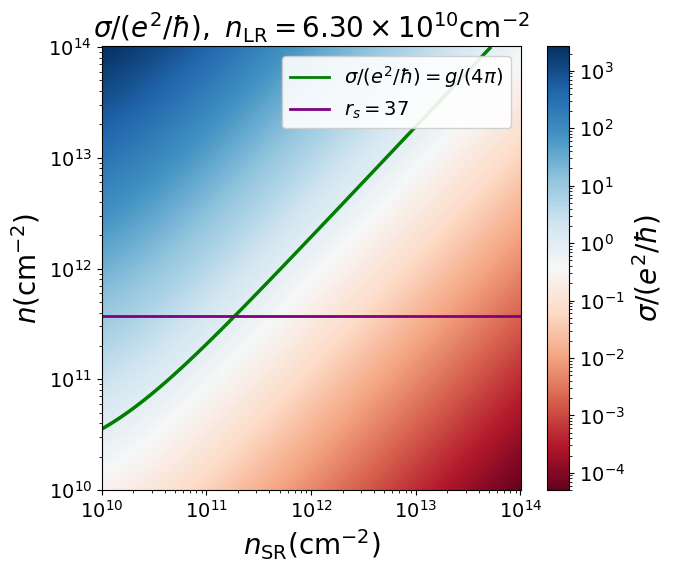}

    \vspace{2pt}

    % --- Second row ---
    \includegraphics[width=0.32\columnwidth]{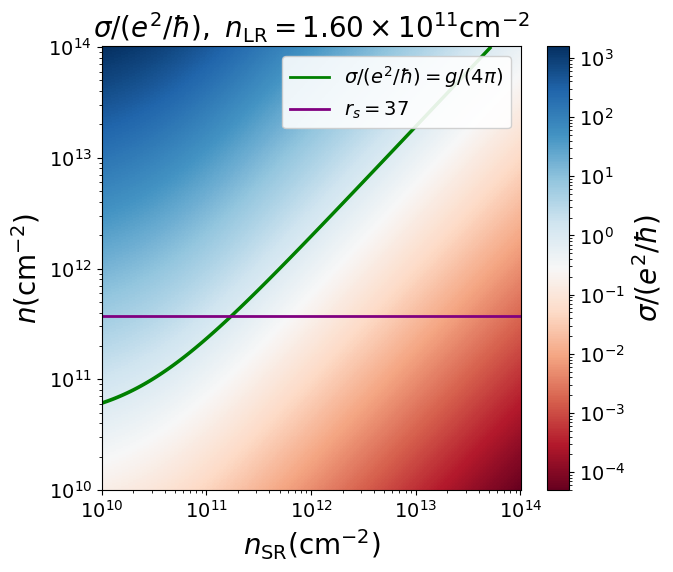}
    \includegraphics[width=0.32\columnwidth]{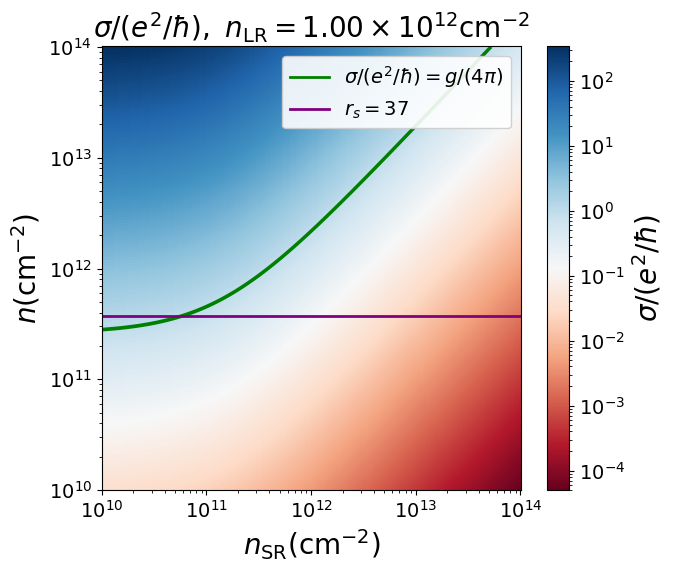}
    \includegraphics[width=0.32\columnwidth]{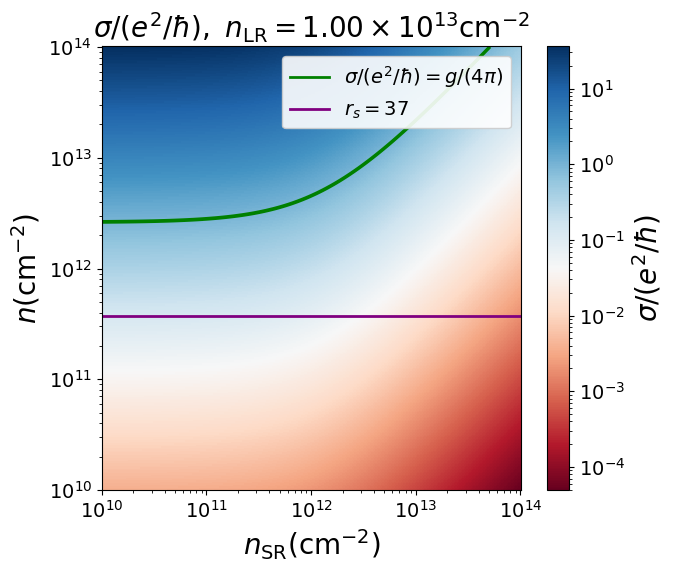}

    \caption{Conductivity maps for various fixed $n_{\rm LR}$,  $\left(\frac{mV_0}{\hbar^2}\right)^2 = 2$, $g = 12$ for varying $n$ and $n_{\rm SR}$}
    \label{fig:nlrfixedvarynandnsr}
\end{figure}

\begin{figure}[h]
    \centering
    % --- First row ---
    \includegraphics[width=0.32\columnwidth]{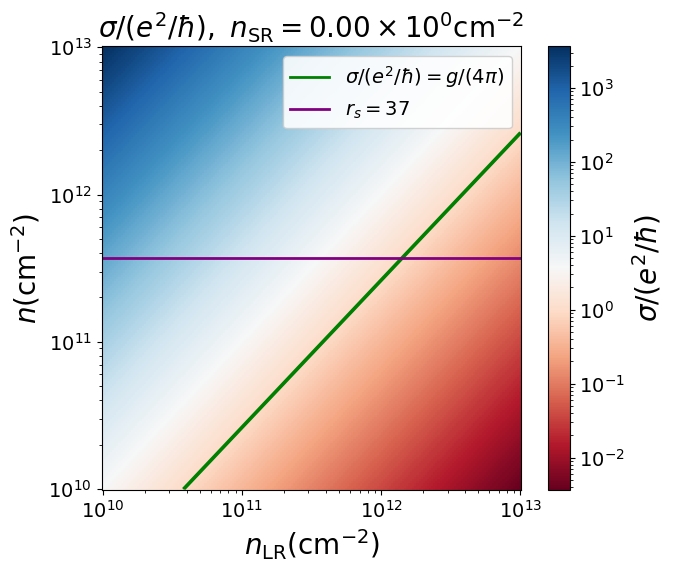}
    \includegraphics[width=0.32\columnwidth]{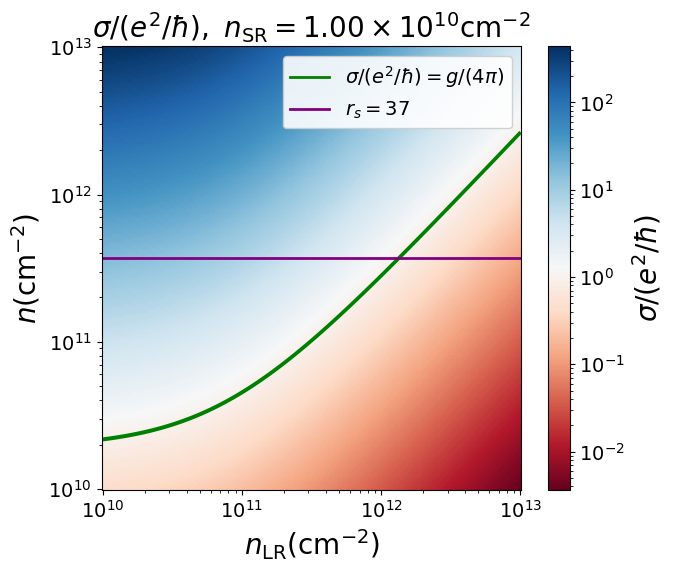}
    \includegraphics[width=0.32\columnwidth]{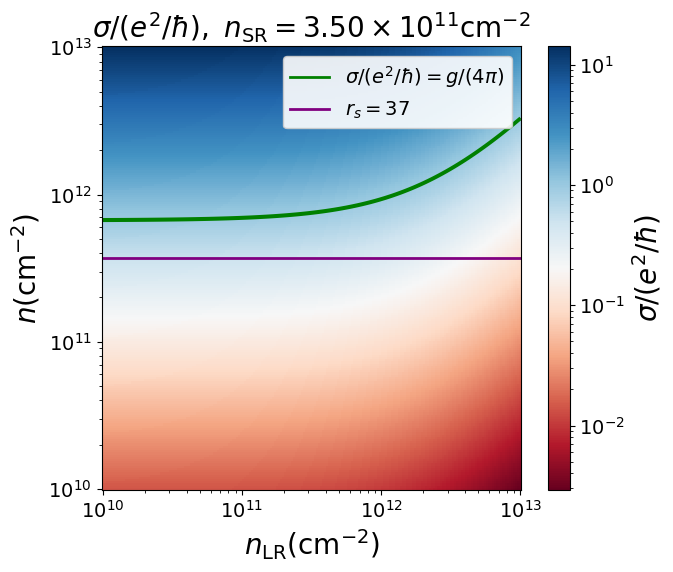}

    \vspace{2pt}

    % --- Second row ---
    \includegraphics[width=0.32\columnwidth]{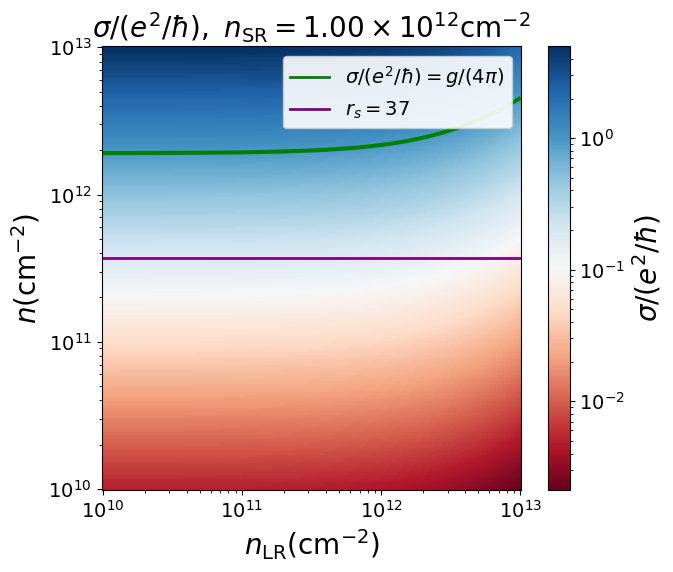}
    \includegraphics[width=0.32\columnwidth]{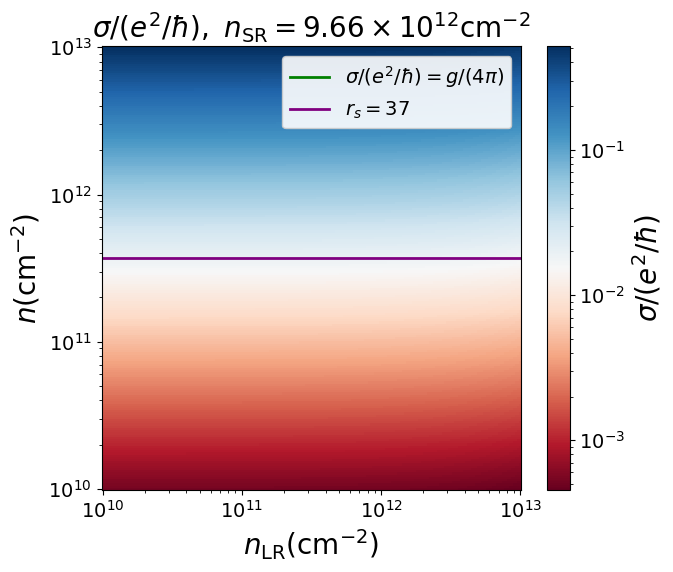}

    \caption{Conductivity maps for various fixed $n_{\rm SR}$,  $\left(\frac{mV_0}{\hbar^2}\right)^2 = 2$, $g = 12$ for varying $n$ and $n_{\rm LR}$}
    \label{fig:nsrfixedvarynandnlr}
\end{figure}

In this section, we plot conductivity as a function of various quantities for $g=12$ (Figures \ref{fig:nlr0parfixedvarynandnsr}, \ref{fig:nlrfixedvarynandnsr}, and \ref{fig:nsrfixedvarynandnlr}).

\begin{figure}[h]
    \centering
    \includegraphics[width=0.8\columnwidth]{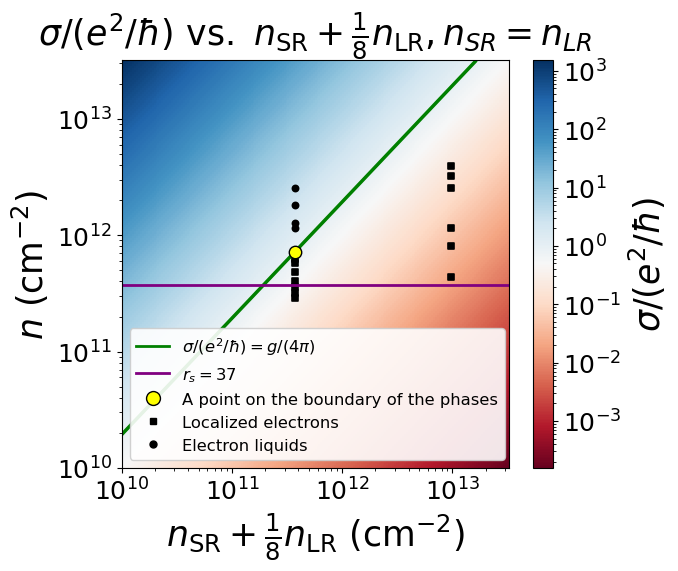}
    \caption{Conductivity as a function of $n_i = n_{\rm SR} + \frac{1}{8} n_{\rm LR}$ (for $n_{\rm SR} = n_{\rm LR}$). The yellow point marks the highest electron density sample that is insulating in the LDD regime, while the black squares and circles represent samples that are electron solids and liquids, respectively.}
    \label{fig:nsrnlr18}
\end{figure}

We note that because our long-range impurities are strongly screened, they behave like short-range disorder in the sense that the scattering rates due to these impurities are electron density independent. We also note that when we compute the conductivity or mobility, because $\left(mV_0/\hbar^2\right)^2 \simeq 2$, the factors next to $n_{\rm{LR}}$ and $n_{\rm{SR}}$ have the ratio $\left(2\pi/g\right)^2/2 \simeq 1/8$. Therefore, plotting the conductivity as a function of $n_{\rm{SR}} +  n_{\rm{LR}}/8$ while overlaying the experimental data points gives Figure \ref{fig:nsrnlr18}. From Figure \ref{fig:nsrnlr18}, we see that an Anderson localization-induced transition for $g=12$ explains the results of the experiment very well.

\section{\large \uppercase\expandafter{\romannumeral1}.5 $g=2$} 

In \cite{ge2025visualizing}, it was pointed out that better agreement with Friedel oscillations is obtained if we set $g=2$. In this section, we analyze the case of $g = 2$ given the transport results of the experiment, and we shall argue that given the value of quoted mobility, $g=2$ is incompatible with the quoted values of disorder.

We recall the original integral to solve in order to compute the scattering rate

\begin{equation}
    \frac{1}{\tau_T} = n_{\rm LR} \frac{2\pi \hbar}{m} \left(\frac{2}{g}\right)^2 2s^2\int_0^1 \text{d}x \, \frac{x^2}{\sqrt{1-x^2} \left(x+s\right)^2}, 
\end{equation}

where $s = q_{TF}/(2k_F)$. The analytic result for the definite integral

\begin{equation}
    \int_0^1 \text{d}x \, \frac{x^2}{\sqrt{1-x^2} \left(x+a\right)^2} \nonumber
\end{equation}

is given in a piecewise fashion (for $a>0$)

\[
\begin{cases}
\displaystyle 
\frac{\pi}{2} - \frac{a}{\sqrt{1-a^2}} \!\left(\ln\!\left(\frac{a+1-\sqrt{1-a^2}}{a+1+\sqrt{1-a^2}}\right) + \ln\!\left(\frac{1+\sqrt{1-a^2}}{1-\sqrt{1-a^2}}\right)\right) - \frac{2}{a} + \frac{1}{a+1} \\[6pt]
\displaystyle 
\quad + \frac{a}{(1-a^2)^{3/2}} \!\left(\ln\!\left(\frac{a+1+\sqrt{1-a^2}}{1+\sqrt{1-a^2}}\right) - \ln\!\left(\frac{1+a-\sqrt{1-a^2}}{1-\sqrt{1-a^2}}\right)\right) \\[6pt]
\displaystyle 
\quad + \frac{a}{1-a^2} \!\left(\frac{1}{1+\sqrt{1-a^2}} + \frac{1}{1-\sqrt{1-a^2}} - \frac{1}{1+a+\sqrt{1-a^2}} - \frac{1}{1+a-\sqrt{1-a^2}}\right), & a < 1 \\[12pt]

\displaystyle 
\frac{\pi}{2} - \frac{4}{3}, & a = 1 \\[12pt]

\displaystyle 
\frac{\pi}{2} - \frac{2a}{\sqrt{a^2-1}} \left(\tan^{-1} \!\left(\frac{a+1}{a^2-1}\right) - \tan^{-1} \!\left(\frac{1}{a^2-1}\right)\right) - \frac{2}{a} + \frac{1}{a+1} \\[6pt]
\displaystyle 
\quad + 2\!\left(\frac{a}{(a^2-1)^{3/2}} \!\left(\tan^{-1} \!\left(\frac{a+1}{a^2-1}\right) - \tan^{-1} \!\left(\frac{1}{a^2-1}\right)\right) + \frac{1}{a^2-1} \!\left(\frac{1}{2} - \frac{1}{a}\right)\right), & a > 1 \\[12pt]
\end{cases}
\]

We set $s = \frac{ge^2 m}{2 \kappa k_F \hbar^2}$, with $k_F = \sqrt{\frac{4 \pi n}{g}}$ for $g = 2$. We have that the total mobility is given by

\begin{equation}
    \mu = \frac{e}{\hbar} \frac{1}{2 \pi \left(\frac{2}{g}\right)^2 \int_0^1 \text{d}x \, \frac{2 s^2 x^2}{\sqrt{1-x^2} \left(x+s\right)^2} n_{\rm LR} \, + \, n_{\rm SR}\left(\frac{m V_0}{\hbar^2}\right)^2}.
\end{equation}

In a similar vein to what we have done previously, we vary the dimensionless $\left(\frac{m V_0}{\hbar^2}\right)^2$ for fixed electron densities $n$ and relative permittivity $\kappa$ to see which value of $\left(\frac{m V_0}{\hbar^2}\right)^2$, for their quoted disorder parameters yields their quoted mobility of $2000 \, \text{cm}^2 \, \text{V}^{-1} \, \text{s}^{-1}$.

\begin{figure}[htbp]
    \centering
    % --- Row 1 ---
    \includegraphics[width=0.32\columnwidth]{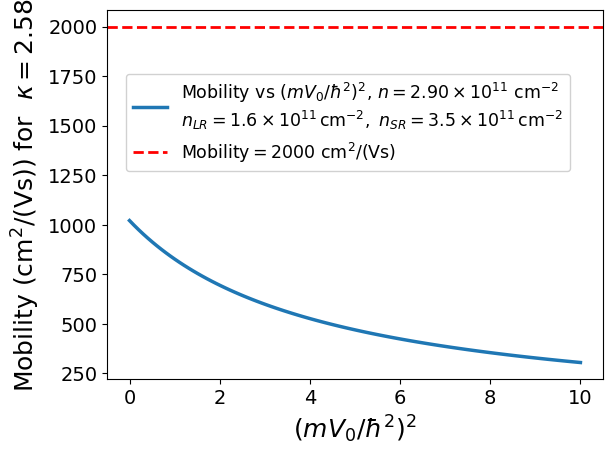}
    \includegraphics[width=0.32\columnwidth]{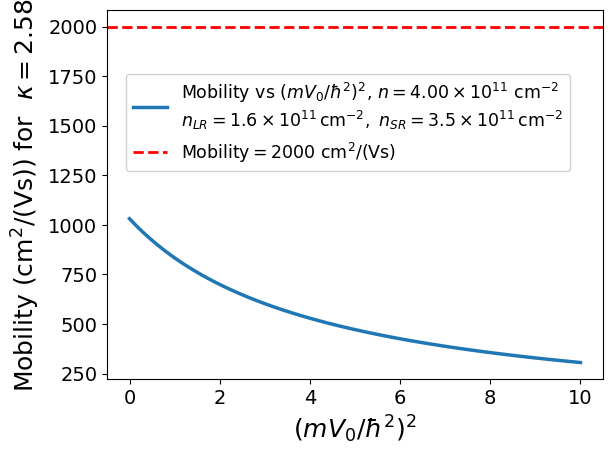}
    \includegraphics[width=0.32\columnwidth]{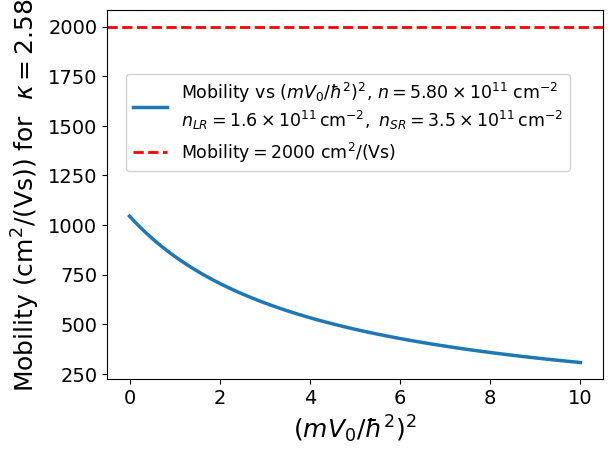}

    \vspace{2pt}

    % --- Row 2 ---
    \includegraphics[width=0.32\columnwidth]{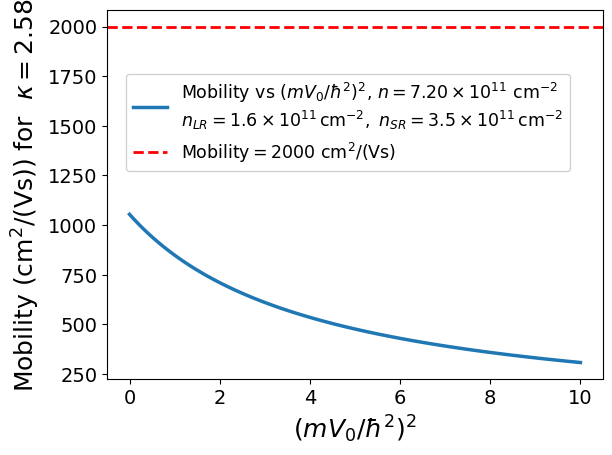}
    \includegraphics[width=0.32\columnwidth]{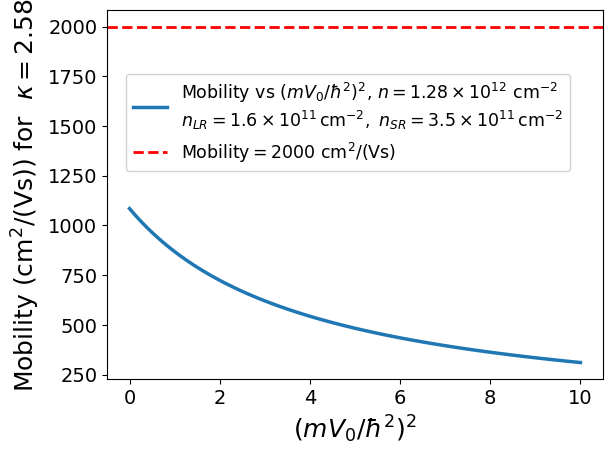}
    \includegraphics[width=0.32\columnwidth]{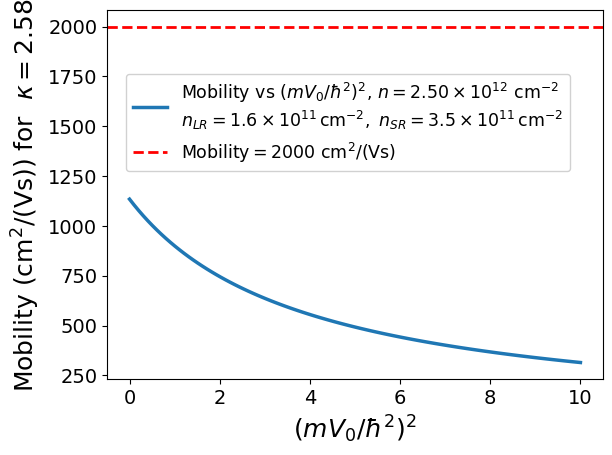}

    \vspace{2pt}

    % --- Row 3 ---
    \includegraphics[width=0.32\columnwidth]{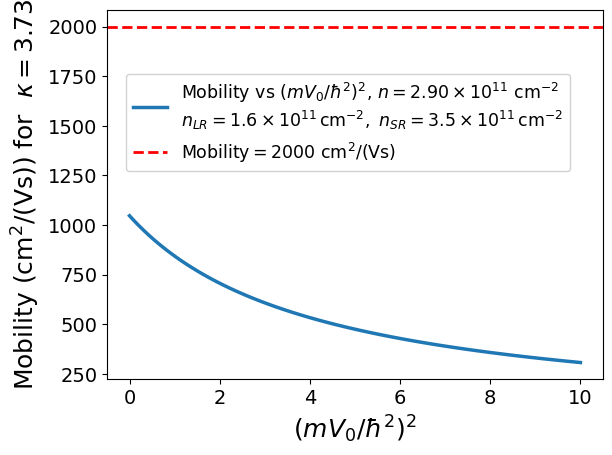}
    \includegraphics[width=0.32\columnwidth]{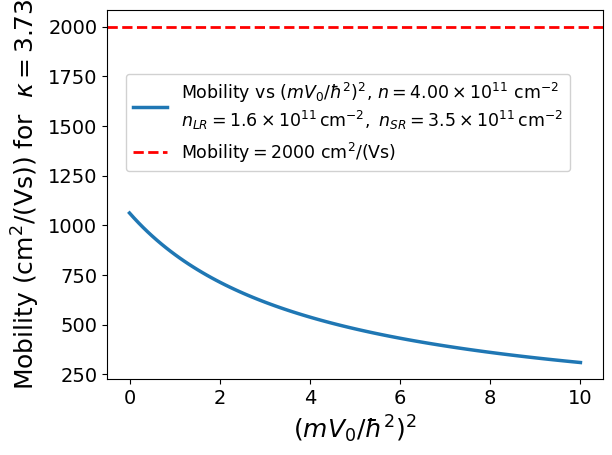}
    \includegraphics[width=0.32\columnwidth]{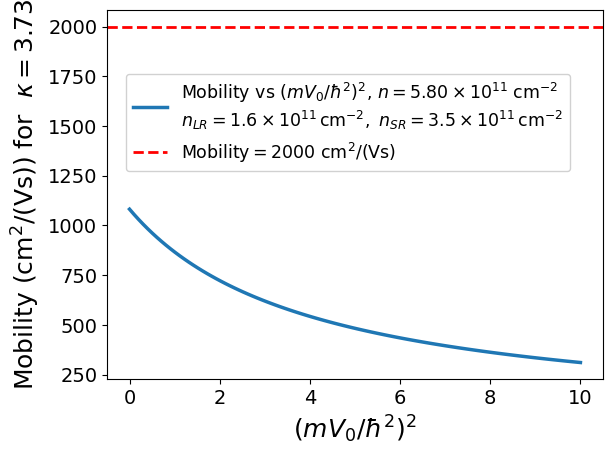}

    \vspace{2pt}

    % --- Row 4 ---
    \includegraphics[width=0.32\columnwidth]{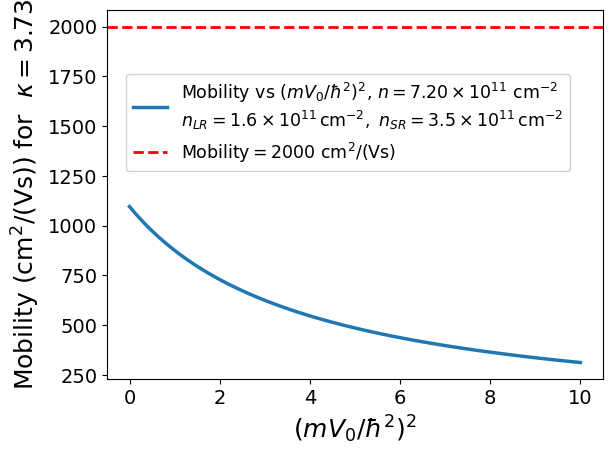}
    \includegraphics[width=0.32\columnwidth]{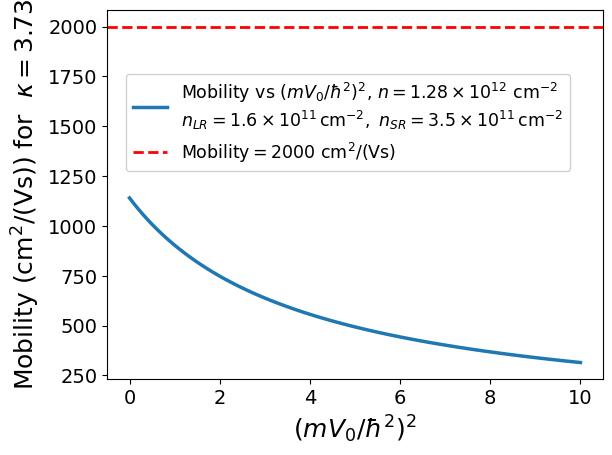}
    \includegraphics[width=0.32\columnwidth]{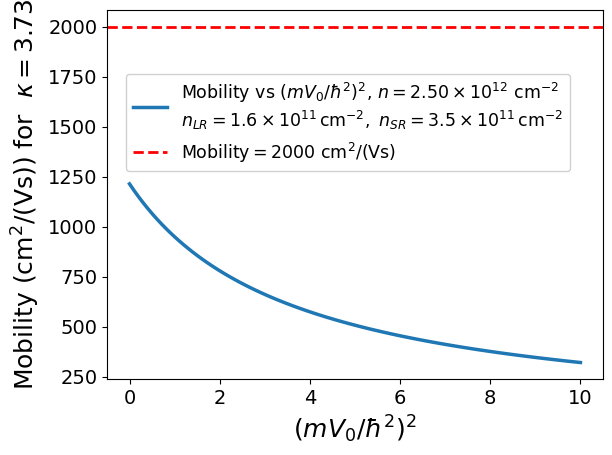}

    \caption{Mobility against $\left(m V_0/\hbar^2\right)^2$ plots for varying $n$ and $\kappa$ for $g = 2$: Low disorder case.}
    \label{fig:mobility_grid_ldd}
\end{figure}

For their low disorder case, $n_{\rm SR} = 3.5 \times 10^{11} \text{cm}^{-2}, \, n_{\rm LR} = 1.6 \times 10^{11} \text{cm}^{-2}$ and we find that for all the $n$ and $\kappa$ they probe, any non-negative $\left(\frac{m V_0}{\hbar^2}\right)^2$ gives $\mu < 2000 \, \text{cm}^2 \, \text{V}^{-1} \, \text{s}^{-1}$ (Figure \ref{fig:mobility_grid_ldd})

\begin{figure}[htbp]
    \centering
    % --- Row 1 ---
    \includegraphics[width=0.32\columnwidth]{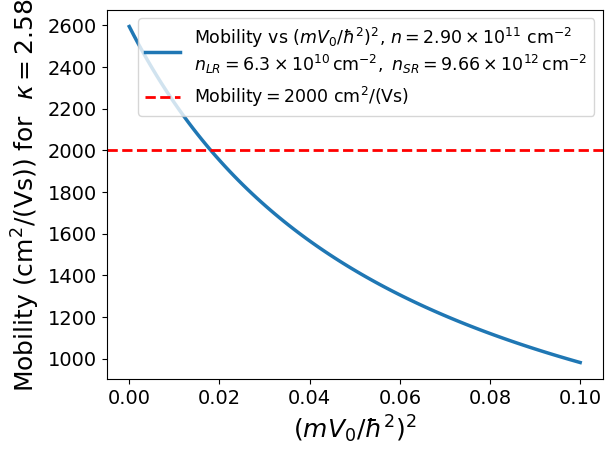}
    \includegraphics[width=0.32\columnwidth]{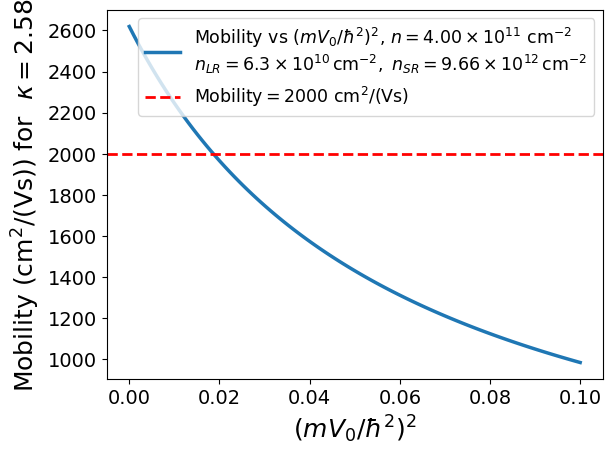}
    \includegraphics[width=0.32\columnwidth]{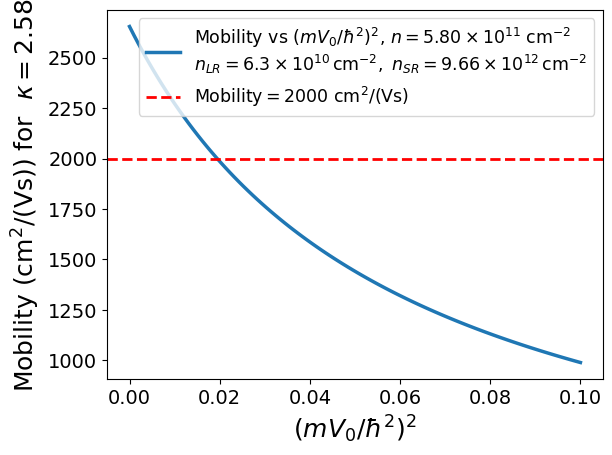}

    \vspace{2pt}

    % --- Row 2 ---
    \includegraphics[width=0.32\columnwidth]{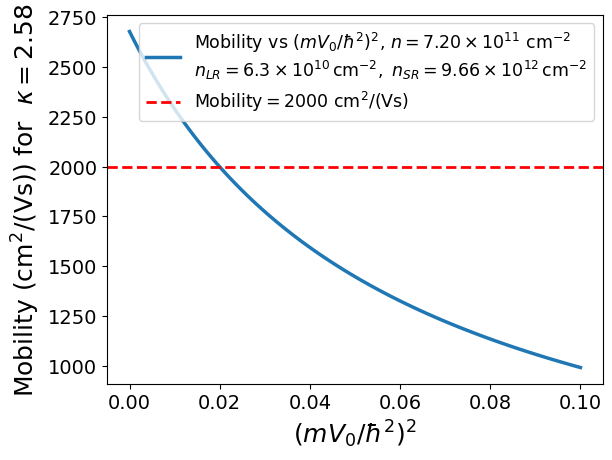}
    \includegraphics[width=0.32\columnwidth]{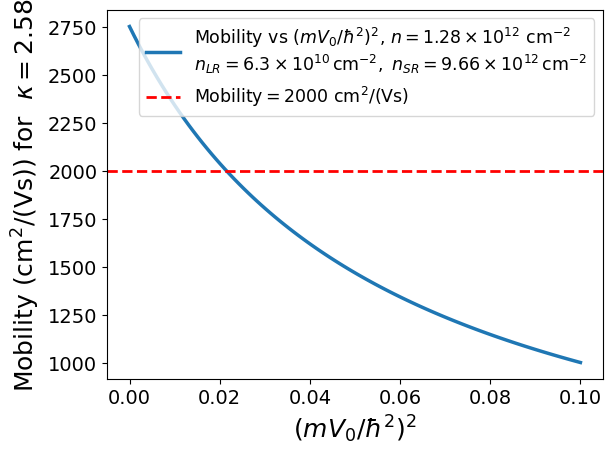}
    \includegraphics[width=0.32\columnwidth]{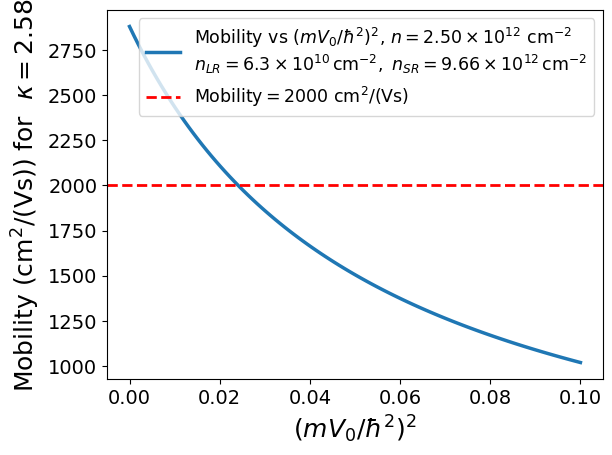}

    \vspace{2pt}

    % --- Row 3 ---
    \includegraphics[width=0.32\columnwidth]{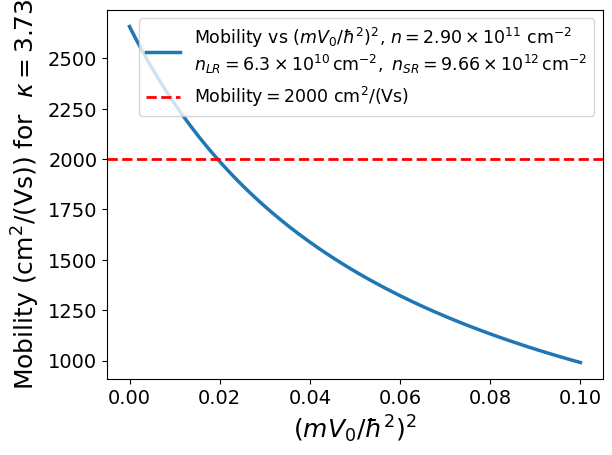}
    \includegraphics[width=0.32\columnwidth]{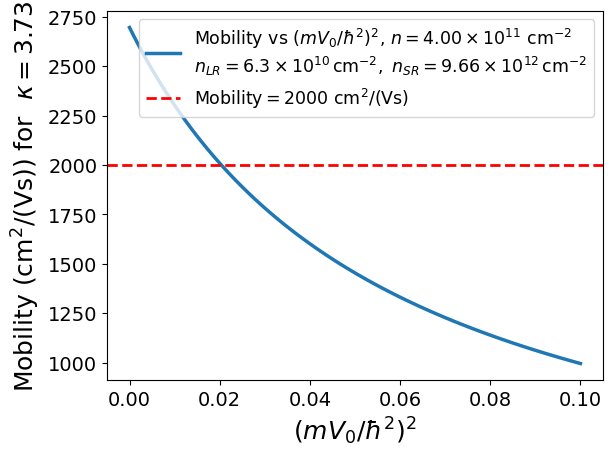}
    \includegraphics[width=0.32\columnwidth]{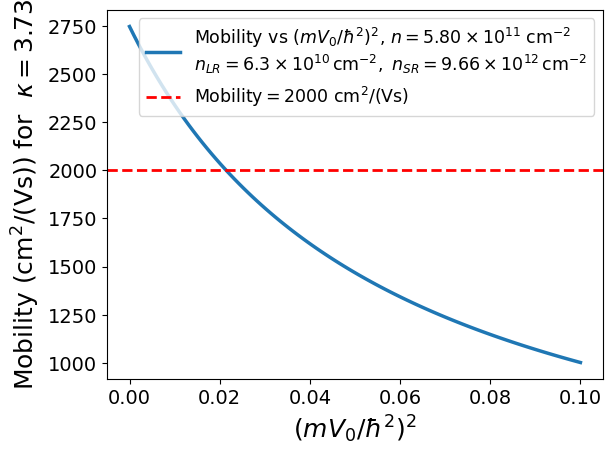}

    \vspace{2pt}

    % --- Row 4 ---
    \includegraphics[width=0.32\columnwidth]{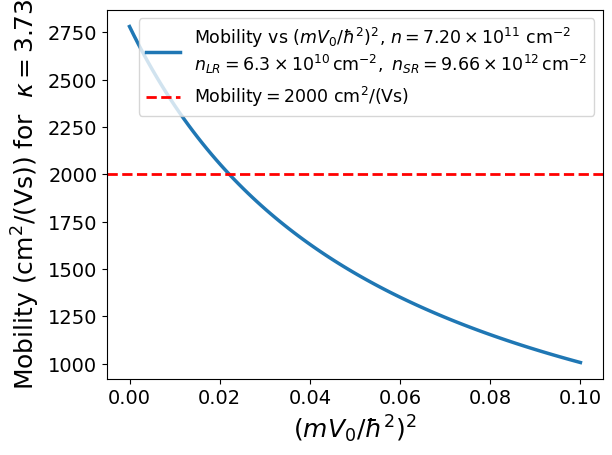}
    \includegraphics[width=0.32\columnwidth]{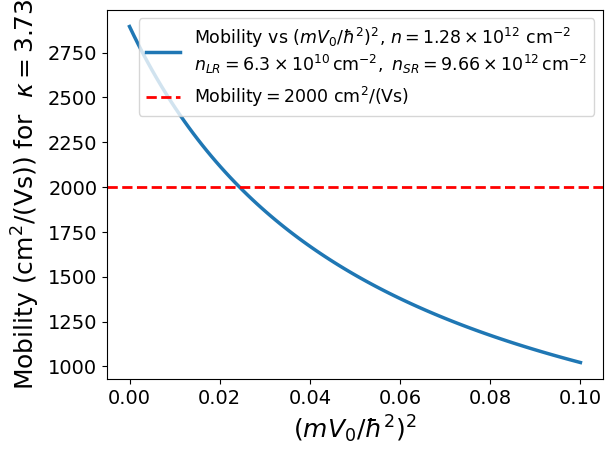}
    \includegraphics[width=0.32\columnwidth]{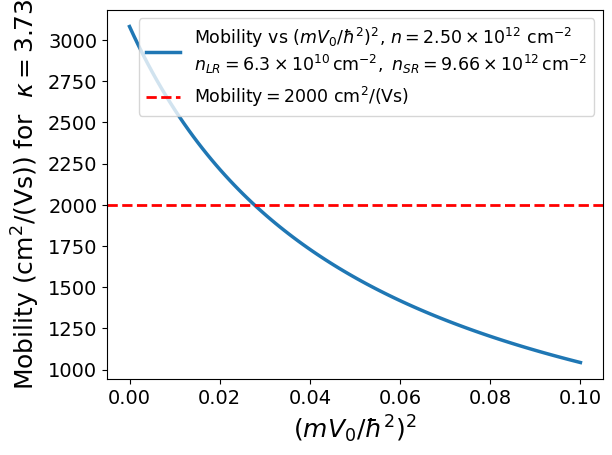}

    \caption{Mobility against $\left(m V_0/\hbar^2\right)^2$ plots for varying $n$ and $\kappa$ for $g = 2$: High disorder case.}
    \label{fig:mobility_grid_hdd}
\end{figure}

If, however, we move to their high disorder case, for which $n_{\rm SR}$ has increased by an order of magnitude ($n_{\rm SR} = 9.66 \times 10^{12} \text{cm}^{-2}$), but $n_{\rm LR}$
has reduced by (roughly) half ($n_{\rm LR} = 6.3 \times 10^{10} \text{cm}^{-2}$), to get $2000 \, \text{cm}^2 \, \text{V}^{-1} \, \text{s}^{-1}$, we need to have a $\left(\frac{m V_0}{\hbar^2}\right)^2 \sim 0.02$, 2 orders of magnitude less than our value of 2 for $g = 12$ (Figure \ref{fig:mobility_grid_hdd}). If we are to go with $g = 2$, it seems like the quoted mobility can only be possible for the high disorder case, which seems incorrect.

Therefore, we say that $g = 2$ is inconsistent with the quoted mobility and disorder content of the system.

\section{\large \uppercase\expandafter{\romannumeral1}.6 Analyzing the unpublished transport data} \label{sec:unpublished}

The authors of \cite{ge2025visualizing} were kind to share unpublished transport data for a very similar sample to the one used in the STM experiment \cite{WangKimPrivateComm}. Analysis of this data sheds a lot of light on the system, and supports the claim that the Metal-insulator transition is indeed an Anderson localization induced transition for a sample with $g=12$.

\subsection{\large \uppercase\expandafter{\romannumeral1}.6.1 Mobility}

A key feature of the experimental Mobility versus electron density curve is that the mobility does not change upon varying the electron density until a low enough electron density is reached. For the lowest temperature recorded (2K) and the highest charge density ($6 \times 10^{12} \text{ cm}^{-2}$) the mobility of this sample is $\sim 3000 \text{cm}^2 \, \text{V}^{-1} \, \text{s}^{-1}$. Until we reach $n \sim 5 \times 10^{11} \text{ cm}^{-2}$, this remains the mobility of the sample, and below this density, the mobility drops sharply, consistent with a metal-insulator transition.

\begin{figure}[h!]
    \centering
    % --- Row 1 ---
    \includegraphics[width=0.48\columnwidth]{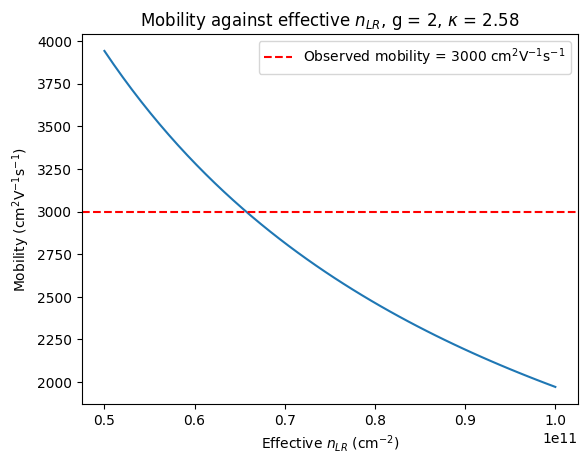}
    \hfill
    \includegraphics[width=0.48\columnwidth]{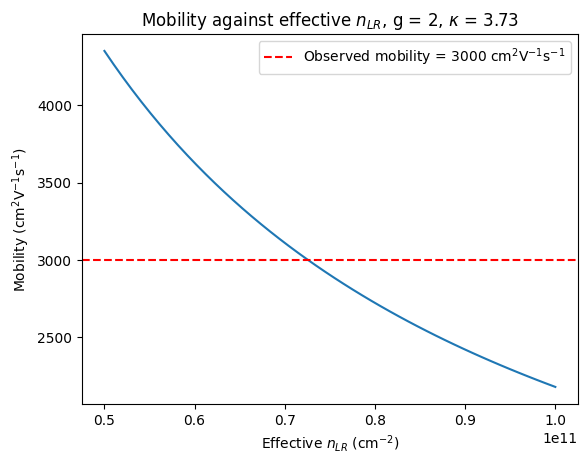}

    \vspace{2pt}

    % --- Row 2 ---
    \includegraphics[width=0.48\columnwidth]{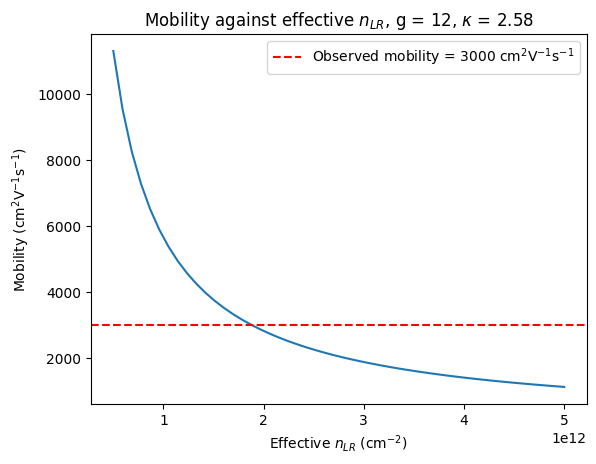}
    \hfill
    \includegraphics[width=0.48\columnwidth]{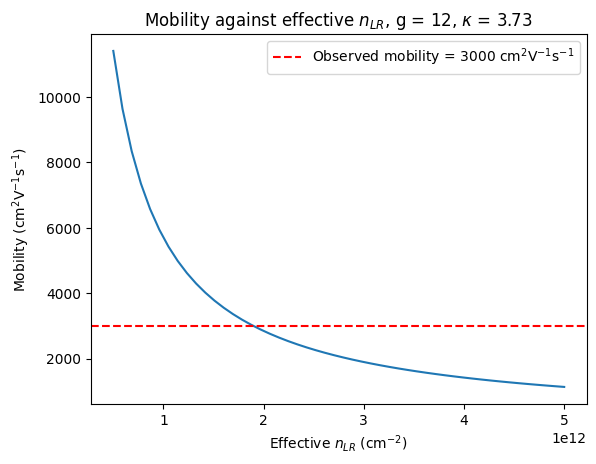}

    \caption{Mobility against $n_i$ (effective $n_{\rm LR}$) for $n = 6 \times 10^{12} \, \text{cm}^{-2}$ for varying $g$ and $\kappa$.}
    \label{fig:mobilityunpublished}
\end{figure}

As we have seen previously, the charged disorder is strongly screened. In this background, we shall treat the system assuming it only has ``effective" strongly screened charged disorder, each with charge $e$. This ``effective" charged disorder has a density $n_i$, which we shall determine using the measurement of mobility. We try to estimate the disorder content $n_i$ by plotting $T = 0$ mobility against $n_i$ for $g= 2, 12$, $\kappa = 2.58, 3.73$ and $n = 6 \times 10^{12} \, \text{cm}^{-2}$ to see what value of $n_i$ yields our required value of $3000 \, \text{cm}^2 \, \text{V}^{-1} \, \text{s}^{-1}$. From Figure \ref{fig:mobilityunpublished}, it can be seen that $n_i \simeq 2 \times 10^{12} \, \text{cm}^{-2}$ yields our required mobility for $g = 12$ and 
$n_i \simeq 7 \times 10^{10} \, \text{cm}^{-2}$ yields our required mobility for $g = 2$. 

We note that $n_i$ should roughly correspond to $n_{\rm LR} + \left(\frac{g}{2 \pi}\right)^2 \left(\frac{m V_0}{\hbar^2}\right)^2 n_{\rm SR}$ if this data is consistent with our previous analysis. For $g = 12$, where we previously obtained $\left(\frac{m V_0}{\hbar^2}\right)^2 = 2$, and given the experimentalists' data for disorder content, $n_{\rm LR} + \left(\frac{g}{2 \pi}\right)^2 \left(\frac{m V_0}{\hbar^2}\right)^2 n_{\rm SR} \simeq 3 \times 10^{12} \text{cm}^{-2}$, which differs from our new result by a factor of $2/3$, and this is precisely because in obtaining this result, we have chosen $\mu = 3000 \, \text{cm}^2 \, \text{V}^{-1} \, \text{s}^{-1}$, as opposed to $2000 \, \text{cm}^2 \, \text{V}^{-1} \, \text{s}^{-1}$, which we did before. For $g = 2$, we do not have a good baseline to compare our new result to, however, we note that their reported value of charged disorder $n_i \simeq 7 \times 10^{10} \, \text{cm}^{-2} < n_{\rm{LR}}$, even though $n_i$ should also carry within it positive contribution from short-ranged disorder in addition to charged, long-ranged disorder.

\subsection{\large \uppercase\expandafter{\romannumeral1}.6.2 IRM criterion}

With the values of $n_i$ we have obtained, we now see at what charged carrier density, given constant disorder density, we meet the IRM criterion. 

\begin{figure}[ht]
    \centering
    % --- Row 1 ---
    \includegraphics[width=0.48\columnwidth]{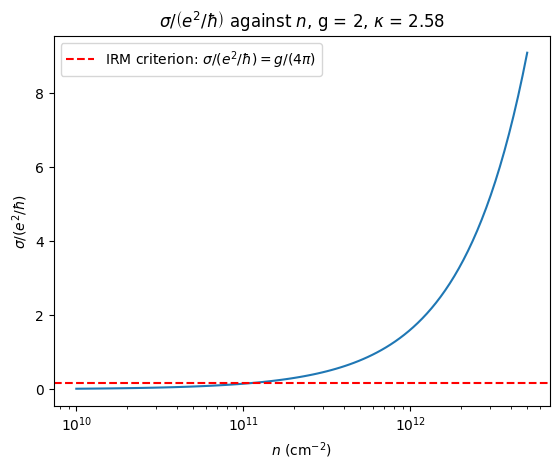}
    \hfill
    \includegraphics[width=0.48\columnwidth]{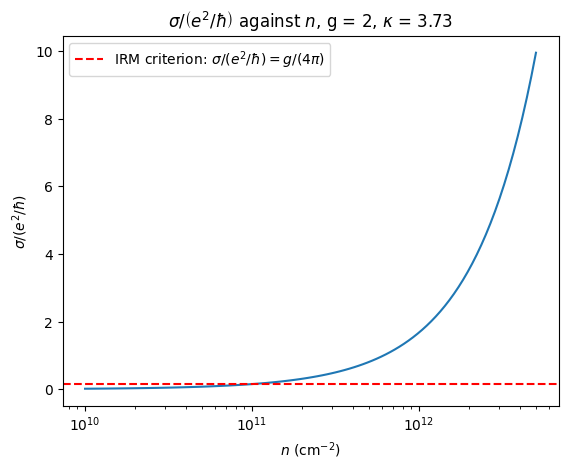}

    \vspace{2pt}

    % --- Row 2 ---
    \includegraphics[width=0.48\columnwidth]{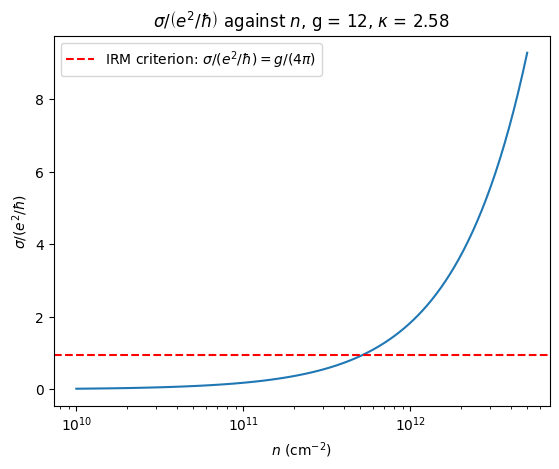}
    \hfill
    \includegraphics[width=0.48\columnwidth]{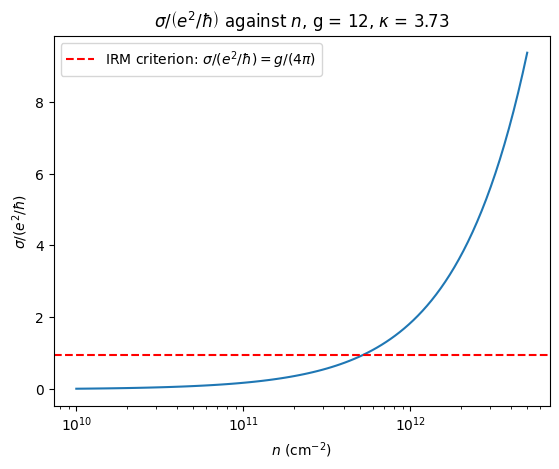}

    \caption{Conductivity against $n$ evaluated with the unpublished data.}
    \label{fig:irmunpublished}
\end{figure}

We plot the conductivity against $n$ for $g = 2, 12$, with their corresponding values of $n_i$ and see that localization is obtained for $n \lesssim 5 \times 10^{11} \, \text{cm}^{-2}$ for $g = 12$ and for $n \lesssim 1 \times 10^{11} \, \text{cm}^{-2}$ for $g = 2$ (Figure \ref{fig:irmunpublished}). As before, the value obtained for $g = 12$ differs from our original value by a factor of roughly $2/3$, which can be explained by the same reason as before.

We note that from their mobility data, the mobility at $T = 2$ K starts dropping sharply at $n \sim 5 \times 10^{11} \, \text{cm}^{-2}$, which is consistent with the IRM criterion's result for $g=12$.

\subsection{\large \uppercase\expandafter{\romannumeral1}.6.3 Temperature dependence of metallic resistivity}

We also note from their data that for $n$ in the metallic regime, we have that the resistivity $\rho$ increases linearly with the temperature $T$ for $T \ll T_F$. This can be explained by the following equation in \cite{DasSarma2015Screening}

\begin{equation} \label{eq:lowTrho}
    \rho = \rho_0 \left(1 + \frac{2x}{1+x} \frac{T}{T_F} + 2.646 \left(\frac{x}{1+x}\right)^2 \left(\frac{T}{T_F}\right)^{3/2} + \mathcal{O} \left(\frac{T}{T_F}\right)^{2} \right).
\end{equation}

\begin{table}[h!] 
\centering
\begin{tabular}{|c|c|c|}
\hline
\textbf{Density} ($10^{11}\,\text{cm}^{-2}$) & \textbf{High T Slope} ($\Omega\,\text{K}^{-1}$) & \textbf{Low T Slope} (k$\Omega\,\text{K}^{-1}$) \\ \hline
14.71 & 0.268 & 0.075 \\ \hline
16.35 & 0.245 & 0.072 \\ \hline
17.98 & 0.235 & 0.072\\ \hline
19.62 & 0.225 & 0.069\\ \hline
21.25 & 0.216 & 0.067\\ \hline
26.16 & 0.200 & 0.059\\ \hline
29.43 & 0.192 & 0.056\\ \hline
31.06 & 0.185 & 0.053\\ \hline
37.60 & 0.166 & -\\ \hline
45.78 & 0.154 & -\\ \hline
65.40 & 0.144 & -\\ \hline
\end{tabular}
\caption{Slope of the line at high and low $T$ as a function of $n$. The low $T$ regime is when disorder contributions to the resistivity are dominant, and the high $T$ regime is when phonon contributions are dominant.}
\label{tab:slopeden}
\end{table}

Here, $x = q_{TF}/\left(2k_F\right)$ and $\rho_0$ is the $T = 0$ resistivity. From the experimental data, we, very crudely, tabulate the slope of the $R(T)-T$ graph obtained by the experimentalists, for small $T$ in Table \ref{tab:slopeden}.

We could also use the high T version of (\ref{eq:lowTrho}) to see how high $T$ resistivity behaves assuming only charged disorder, however, we note that for the lowest densities we probe, $T_F \sim 10$ K, so in order to use the high $T$ version of (\ref{eq:lowTrho}), we would need $T \gg 10$ K, beyond which phonon effects start to matter. This was observed in experimentalists' unpublished data, where at very high $T$, the resistivity for all densities increases $\propto T$. If we were purely considering scattering from disorder, the resistivity would have had a $\propto 1/T$ behavior. We can explain the experimental data of the resistivity-temperature curve using two separate slopes: one for when $T \ll T_F$, where the disorder has the dominant contribution, and one for high $T$, when the phonon effects dominate. The slopes for both low T and high T are tabulated in Table \ref{tab:slopeden}.

\subsection{\large \uppercase\expandafter{\romannumeral1}.6.4 Valley curves for resistivity}

\begin{figure}
    \centering
    \includegraphics[width=0.4\columnwidth]{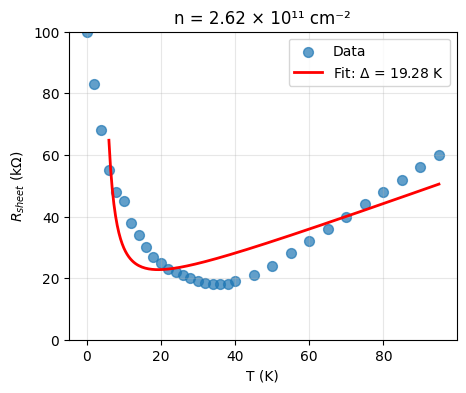}
    \hfill
    \includegraphics[width=0.4\columnwidth]{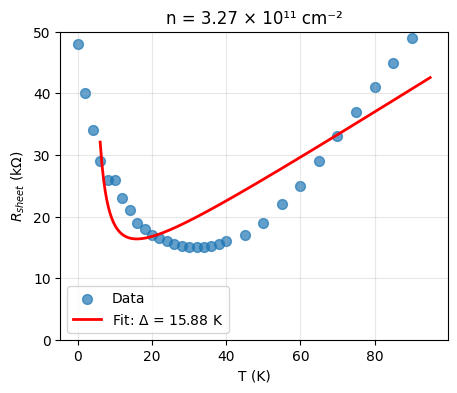}
    \caption{Valley resistivity curves (against $T$) for samples localized at $T = 0$ fitted to $\rho = A T e^{\Delta/T}$.}
    \label{fig:fit}
\end{figure}

Interestingly, we note that for certain values of charge carrier density that are localized at $T = 0$, as we increase the temperature, the resistivity drops to a minimum and then increases linearly at high $T$. We attribute the rise of this ``valley-like" curve at high $T$ to phonons, which give a $\propto T$ resistivity; however, in order to explain the valley, we propose an ``activation" term in the resistivity that goes like $e^{\Delta/T}$ that when multiplied with the linear in $T$ term gives a curve that has a minimum, much like we see in the data and increases linearly in $T$ for large $T$. Some representative Resistivity-temperature curves for $n$ in this regime are shown as the dotted data points in Figure \ref{fig:fit}.

\begin{figure}
    \centering
    \includegraphics[width=0.5\columnwidth]{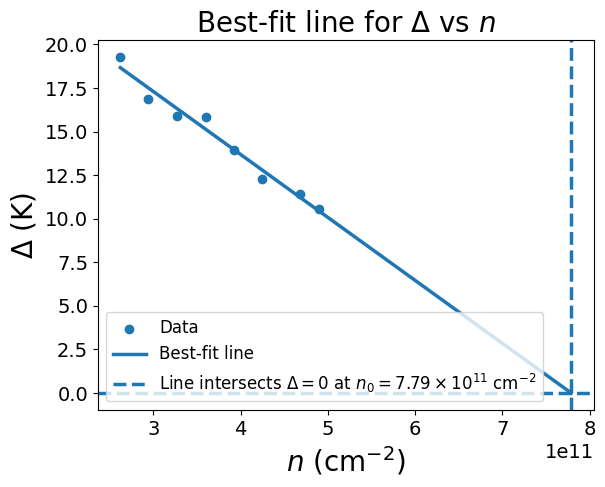}
    \caption{$\Delta$ for various values of $n$ plotted and fit to a straight line. The value at which the straight line intersects $\Delta = 0$ is another prediction for when the Metal-Insulator transition takes place.}
    \label{fig:straight_line_mit}
\end{figure}

We propose $\rho \simeq A T e^{\frac{\Delta}{T}}$, and we fit the experimental data for various charge carrier densities to find $\Delta$ in Figure \ref{fig:fit}. The fits are not very accurate because we totally ignore the contribution of disorder to resistivity; however, this idea is enough to get a rough physical picture. We plot the values of $\Delta$ for various $n$ in the insulating regime. In fact, when we plot the values of $\Delta$ against $n$, we find that $\Delta$ indeed decreases with increasing $n$ when we are in the insulating regime. We expect that as the Metal-Insulator transition happens, the value of $\Delta$ becomes 0. When we plot the values of $\Delta$ against $n$, fit them to a straight line and extrapolate the straight line to $\Delta = 0$, we find that the $n$ at which this happens is $n \sim 7.79 \times 10^{11} \text{ cm}^{-2}$ (Figure \ref{fig:straight_line_mit}), which is not far from both the value at which the transition is experimentally observed $\left(\sim 7.1 \times 10^{11} \text{ cm}^{-2}\right)$ and the value at which the ``valleys" stop appearing $\left(\sim 7 \times 10^{11} \text{ cm}^{-2}\right)$.

\subsection{\large \uppercase\expandafter{\romannumeral1}.6.5 T = 0 resistivity}

\begin{figure}[htbp]
    \centering
    % --- Row 1 ---
    \includegraphics[width=0.45\columnwidth]{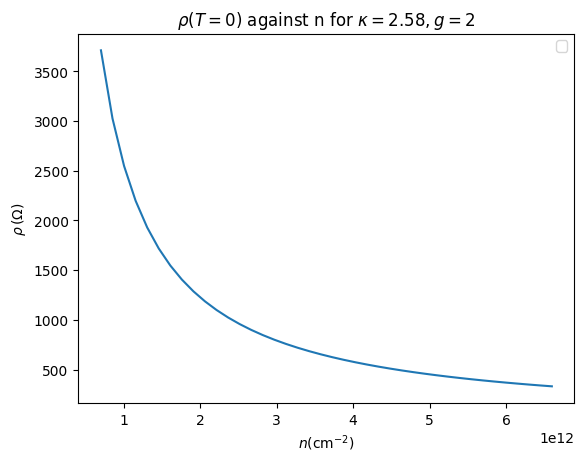}
    \hfill
    \includegraphics[width=0.45\columnwidth]{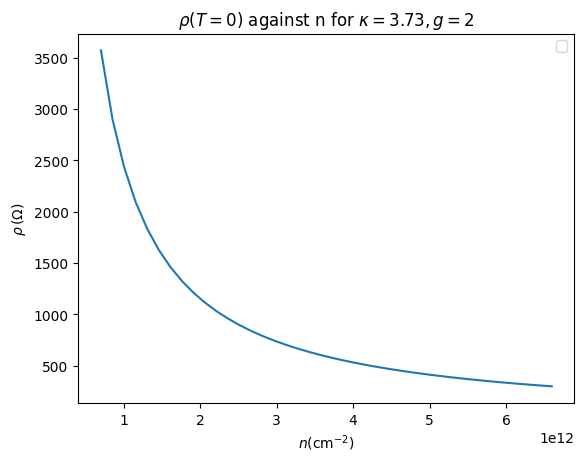}

    \vspace{2pt}

    % --- Row 2 ---
    \includegraphics[width=0.45\columnwidth]{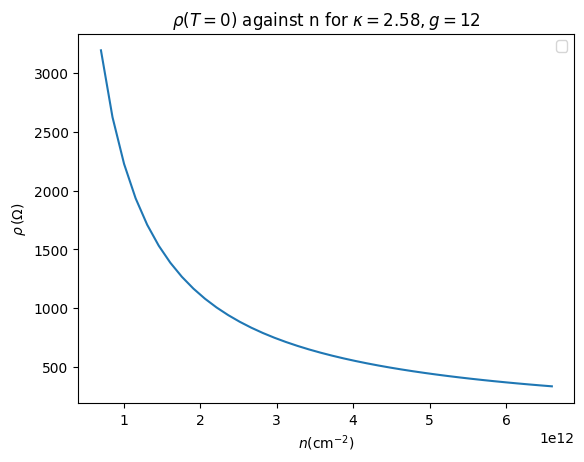}
    \hfill
    \includegraphics[width=0.45\columnwidth]{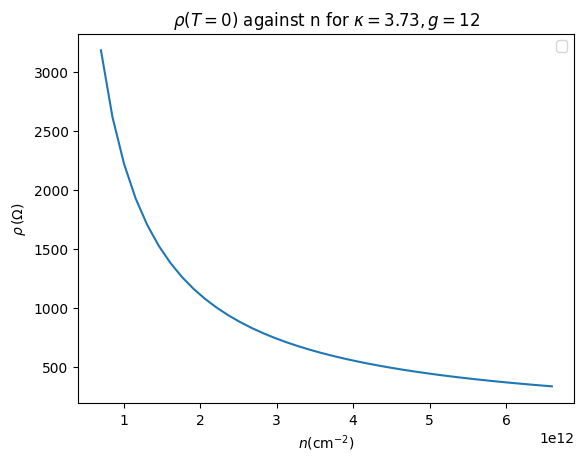}

    \caption{$T = 0$ resistivity against $n$ for $g = 2, 12$ and $\kappa = 2.58, 3.73$.}
    \label{fig:rhoT0}
\end{figure}

We have that the resistivity at $T = 0$ is given by

\begin{equation}
    \rho \, (T=0) = \frac{\hbar}{e^2} \frac{2 \pi \left(\frac{2}{g}\right)^2 \int_0^1 \text{d}x \, \frac{2 s^2 x^2}{\sqrt{1-x^2} \left(x+s\right)^2} n_{\rm LR} \, + \, n_{\rm SR}\left(\frac{m V_0}{\hbar^2}\right)^2}{n},
\end{equation}

where $s = q_{TF}/\left(2 k_F\right)$; however, like before, we treat our system as one with only an effective long-range disorder with density $n_i$. Using this idea and with the measured mobility at $T$ close to 0, we had calculated $n_i$ for both $g = 2$ and $g = 12$. We had that $n_i \, \left(g=2\right) = 7 \times 10^{10} \text{ cm}^{-2}$ and $n_i \, \left(g=12\right) = 2 \times 10^{12} \text{ cm}^{-2}$. We plot, for both values of $g$, each with their respective $n_i$ and both values of $\kappa$ in the experiment the theoretical value of the $T = 0$ resistivity and $q_{TF}/\left(2 k_F\right)$ as a function of the charge carrier density $n$ in Figure \ref{fig:rhoT0}.

\begin{figure}[htbp]
    \centering
    % --- Row 1 ---
    \includegraphics[width=0.45\columnwidth]{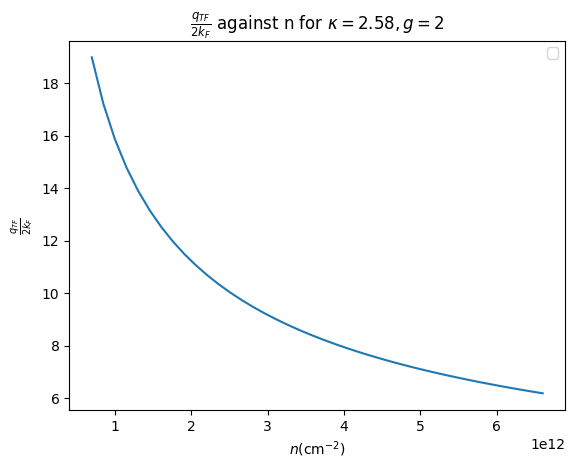}
    \hfill
    \includegraphics[width=0.45\columnwidth]{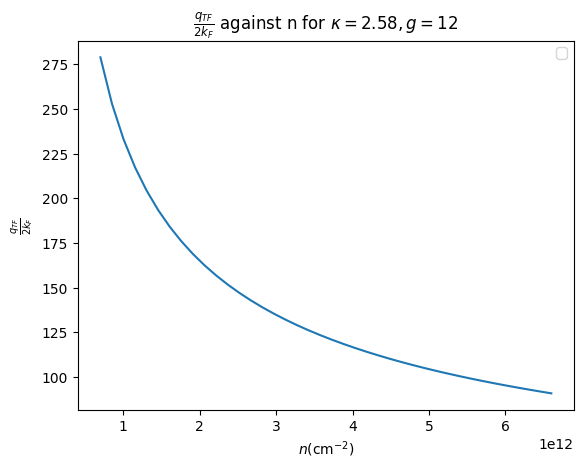}

    \vspace{2pt}

    % --- Row 2 ---
    \includegraphics[width=0.45\columnwidth]{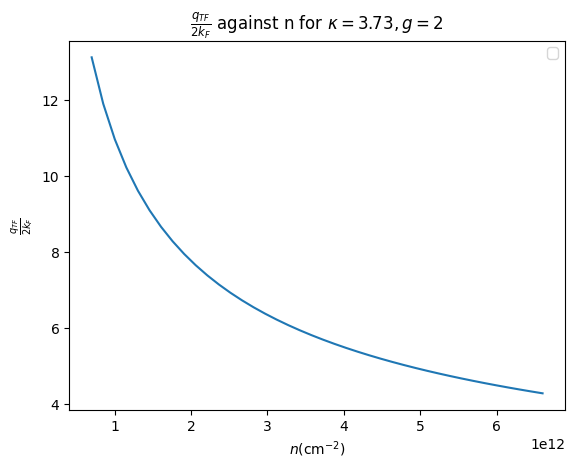}
    \hfill
    \includegraphics[width=0.45\columnwidth]{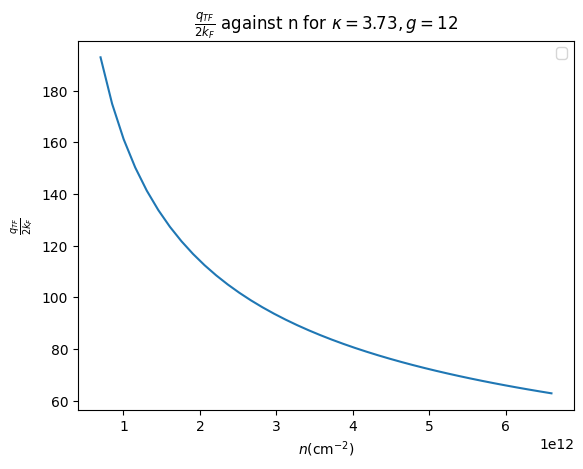}

    \caption{$q_{TF}/\left(2 k_F\right)$ against $n$ for $g = 2, 12$ and $\kappa = 2.58, 3.73$.}
    \label{fig:s}
\end{figure}

From the unpublished transport data, all T = 0 resistivities for $n$ between $\sim 1.5 \times 10^{12} \text{ cm}^{-2}$ and $\sim 6.5 \times 10^{12} \text{ cm}^{-2}$ lie in $0 - 2 \, \text{k} \Omega$ range, which is also roughly what we observe in our theoretical calculation with the values of $n_i$ from above (Figure \ref{fig:rhoT0}). Plotting $s = q_{TF}/\left(2 k_F\right)$ as a function of $n$, we find that for $g = 12$, we have strong screening as $s \gg 1$ (Figure \ref{fig:s}) within the range of densities probed. Even for $g = 2$, we have that $s$ is always larger than $\sim 4$ within the range probed. 

We recall from the unpublished experimental data that the mobility is a constant until it suddenly starts dropping after a metal-insulator transition. This fact is not surprising given that for both values of $g$, screening is sufficiently strong that the scattering time is not dependent on $n$. Another point of note is that the value of $T = 0$ $\rho$ at which the IRM criterion is met is $\rho = \left(\hbar/e^2\right) \left(4 \pi/g\right)$. For $g = 2$, this is $\sim 25 \text{k}\Omega$ and for $g = 12$, this is $\sim 4.3 \text{k} \Omega$. If we were to look at the resistivity against temperature curves for $n$ less than that given by the IRM criterion, we would expect the ``valley" curves as explained in a previous subsection - this is another way to find from the experimental data where the metal-insulator transition happens. From the plots shared by the experimentalists, we find that the upon increasing charge carrier density, the ``valley" curves end at around $\sim 7 \times 10^{11} \, \text{cm}^{-2}$, where the $T = 0$ resistivity is $\sim 5 \text{k} \Omega$. This value agrees a lot more with the value given by the IRM criterion for $g = 12$ than it does with that for $g = 2$. This is another point in favor of the argument that $g = 12$ is in play within our system.

\section{\large \uppercase\expandafter{\romannumeral1}.7 Holes as the charge carriers}

We now discuss transport in a similar system where a Metal-Insulator transition is reported for another sample of Bilayer $\rm{MoSe}_2$, only this time the charge carriers are holes of mass $m = 1.26 m_e$ \cite{xiang2025imaging}. We still assume that the valley degeneracy $g_v = 6$. Given that the experimentalists for this experiment do not mention neutral disorder, we treat all disorder as charged disorder, and because $q_{TF}/\left(2k_F\right) \gg 1$, we have that the mobility, like before, is given by

\begin{equation}
    \mu = \frac{e \tau_t}{m} = \frac{e}{\hbar} \frac{1}{n_{\rm LR} \left(\frac{2\pi}{g}\right)^2}.
\end{equation}

Conductivity, as we know, is given by

\begin{equation}
    \sigma = n e \mu = \frac{e^2}{\hbar} \frac{n}{n_{\rm LR} \left(\frac{2\pi}{g}\right)^2}.
\end{equation}

As the transition happens at around $r_S \sim 20$, we hypothesize that this transition is not a Wigner crystal transition, but instead an Anderson-localization induced transition. We observe from the experimental data that the transition happens at around $n \sim 6.1 \times 10^{12} \text{ cm}^{-2}$. Using this as the density at which the IRM criterion is met,

\begin{equation}
    \frac{g}{4 \pi} = \frac{6.1 \times 10^{12} \text{ cm}^{-2}}{n_{\rm LR} \left(\frac{2\pi}{g}\right)^2}.
\end{equation}

Therefore, $n_{\rm LR} \sim 2.3 \times 10^{13} \text{ cm}^{-2}$. This means that $\mu \simeq 240 \text{ cm}^2 \text{ V}^{-1} \text{ s}^{-1}$. Experimentally, the mobility of this sample is low enough that it could not be measured, but this value is certainly within what the experimentalists deem feasible for the system \cite{WangPrivateComm}. We plot the conductivity as a function of $n$ and $n_{\rm{LR}}$ and overlay it with experimental data points obtained by calculating $n_{\rm{LR}}$ using this mobility (Figures \ref{fig:transport_figure_2}c and \ref{fig:transport_figure_2}d).

\subsection{\Large \uppercase\expandafter{\romannumeral2}. Phase diagram from exact diagonalization}
In this section, we present our exact diagonalization (ED) study of the disorder effects on 2D interacting electrons. Section {\red \uppercase\expandafter{\romannumeral2}.1} introduces a model describing the disordered interacting electrons sitting on a 2D lattice, solved by ED. In Section {\red \uppercase\expandafter{\romannumeral2}.2}, we analyze the ground state properties of electrons at characteristic parameters through the electron local density and density-density correlation function. In Section {\red \uppercase\expandafter{\romannumeral2}.3}, we compute the phase diagram using quantities extracted from the density and correlation, and analyze the consequences of different model parameters on the phase diagram.

\section{\large \uppercase\expandafter{\romannumeral2}.1 Model} 
The model describing $N_e$ electrons sitting on a 2D periodic rectangular lattice with $N = L_x\times L_y$ sites, where $N_i$ impurities randomly locate on the lattice sites is
\begin{equation}
    H = \sum_{\langle i,j\rangle} t(c_{i}^\dagger c_j + h.c.) + \sum_{i,j=1}^{N} U_{ij} n_i n_j + \sum_{i=1}^N W_i n_i,
\end{equation}
where $\langle i,j\rangle$ represents the nearest neighbor sites, $t$ is the hopping strength, $U_{ij}$ is the electron-electron long-range Coulomb interaction, and $W_i$ is the disorder potential. The Coulomb interaction is 
\begin{equation}
    U_{ij} = \frac{U_0}{D(\mathbf r_i - \mathbf r_j)},
\end{equation}
where
\begin{equation}
    D(\mathbf r)^2 =  \left[\frac{L_x}{\pi} \sin \left(\frac{\pi x}{L_x}\right)\right]^2 + \left[\frac{L_y}{\pi} \sin \left(\frac{\pi y}{L_y}\right)\right]^2.
\end{equation}
And the disorder potential is
\begin{equation}
    W_i = -\sum_{j=1}^{N_i} \frac{W e^{-\lambda D(\mathbf r_i - \mathbf R_j)}}{D(\mathbf r_i - \mathbf R_j) + \delta},
\end{equation}
where $W$ is the disorder strength, and $\lambda$ is a screening parameter. The impurities are randomly located on the lattice at $\mathbf R_j$. The short-distance cut-off is fixed at $\delta=1$ to remove singularities.
$\lambda = 0$ represents the long-range disorder, while for the short-range disorder phase diagram in the main text, we set $\lambda = 4$. 
For $W>0$, the impurities are attractive, and vice versa. 

We introduce dimensionless parameters to characterize different regimes of the system: the relative impurity strength $z = W/t$, which measures the effects of disorder, and the effective Wigner-Seitz radius $r_s$, defined as the electron average spacing divided by the Bohr radius, which measures the electron-electron interaction.
Comparing the model parameter to the standard definition of $r_s$, we have $r_s= [\pi\nu]^{-1/2}/[U/(2t)]$, where $\nu \equiv N_e/N$ is the filling fraction. Another useful dimensionless parameter as a verification for our theory in the strongly interacting regime is the average WC lattice spacing, which can be estimated as $a_W = \nu^{-1/2}$.

The WC transition occurs when the interaction dominates the kinetic energy, with a theoretical estimate around $r_s \sim 38$ for a clean system. 
Experimentally, achieving such a parameter range can be accomplished by tuning the voltage to change the carrier density.
This is inaccessible in our model due to the computational limit, since lowering the density while capturing the many-body properties requires increasing the size of the lattice, and therefore results in an exponentially increasing Hilbert space dimension. However, tuning the interaction or the kinetic energy changes $r_s$ as well, which is usually experimentally infeasible but easily achieved theoretically, just by changing $U/t$. In the following calculations, unless mentioned, we will change the interaction to obtain the phase diagram instead of changing the electron densities. And we always perform our calculations at zero temperature, in other words, for the ground state.

\begin{figure}
    \centering
    \includegraphics[width=\linewidth]{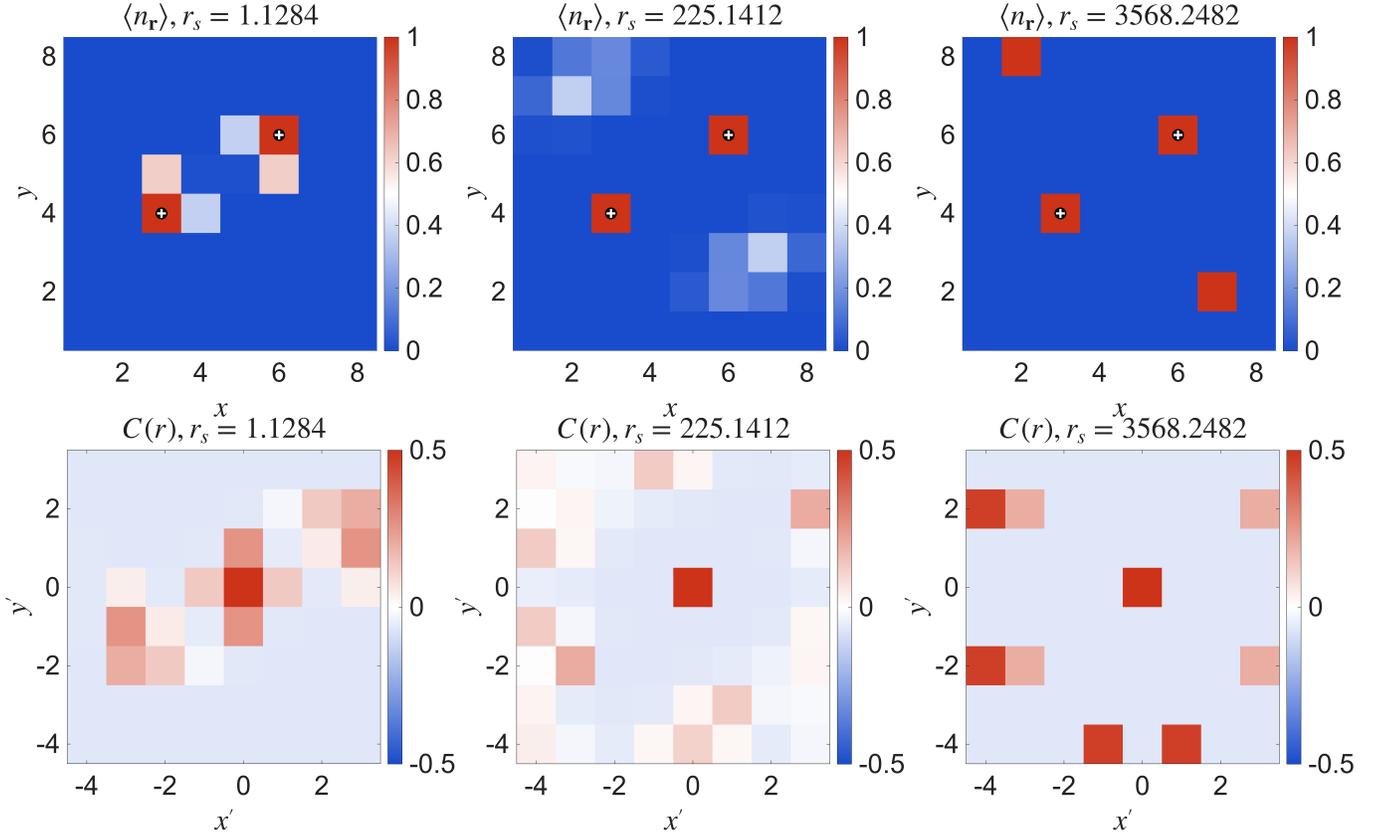}
    \caption{\justifying{Electron local density and correlation function from exact diagonalization. Two attractive disorders are located at $(3,4)$ and $(6,6)$. The disorder strength is fixed at $z\approx 316$. For all figures, $N_e = 4, L_x = L_y = 8$, and periodic boundary conditions are used.}}
    \label{fig:denco}
\end{figure}
\section{\large \uppercase\expandafter{\romannumeral2}.2 Ground state properties}
We display the electron local structure to see whether the physics of Anderson localization and WC are qualitatively captured in the ground state of the Hamiltonian, by computing two quantities using exact diagonalization: the electron local density $\langle n_{\mathbf{r}}\rangle$ and the normalized density-density correlation,
\begin{equation}
    C(\mathbf r) = \frac{1}{1-\nu}\left[\frac{1}{N_e}\sum_{j=1}^N \langle n_{\mathbf r_j+\mathbf r} n_{\mathbf r_j}\rangle - \nu \right].
\end{equation}
Fig.~\ref{fig:denco} shows the $\langle n_{\mathbf{r}}\rangle$ and $C(\mathbf r)$ for the Hamiltonian with two attractive disorders at different interactions $r_s$ and a fixed disorder strength $z\approx 316$. 
When the disorder dominates the interaction, the system is by definition Anderson localized, where the electrons are localized around the disorders, as shown in Fig.~\ref{fig:denco} when $r_s \approx 1.13$. When the interaction dominates the disorder, the system should be a Wigner crystal, where the electrons form a 2D triangular lattice, pinned by the impurities, which is captured in Fig.~\ref{fig:denco}, although only $4$ electrons are used in the calculations. 
In the intermediate regime (Fig.~\ref{fig:denco} where $r_s \approx 225$), part of the electron delocalizes due to the competition between the disorder and interaction, while the others are pinned by the impurities as usual. These qualitative agreements support our method using a small system to capture the general behavior of disorder effects on the interacting electrons.

\section{\large \uppercase\expandafter{\romannumeral2}.3 Phase diagram}
To obtain the phase diagram, we characterize two quantities extracted from the $\langle n_{\mathbf{r}}\rangle$ and $C(\mathbf r)$. One is the many-body inverse participation ratio (MBIPR),
\begin{equation}
    \mathcal{I} = \frac{1}{1-\nu}\left[\frac{1}{N_e}\sum_{j=1}^N \langle n_{\mathbf r_j}\rangle^2 - \nu \right],
\end{equation}
which is a localization measure of electrons, distinguishing the metallic Fermi liquid regime, which is extended, from other localized regimes. For fully localized states, $\mathcal{I} \rightarrow 1$, while for fully extended states, $\mathcal{I} \rightarrow 0$. 
Note that since the localization may be caused by interaction or disorder, corresponding to a pinned WC or Anderson localization, it is necessary to introduce a correlation measure distinguishing these two regimes, i.e., an effective correlation length 
\begin{equation}
    \xi^2 = \frac{\sum r^2 C(r)}{\sum C(r)},
\end{equation}
where the sum runs over the local maxima of $C(r)$. 
We mention that $\xi$ may not directly relate to the physical correlation length in real materials, which is unlikely to be captured in such a small system, but can successfully distinguish different regimes of WC and Anderson localization.
In the Wigner regime, ideally, local maxima of $C(\mathbf r)$ should appear at $\mathbf r_{m} \approx a_W(n \mathbf a_1 + m \mathbf a_2)$, where $\mathbf a_1$ and $ \mathbf a_2$ are lattice unit vectors of the Wigner crystal, and $n,m$ are integers. Since there are only a few electrons in our system, we have $|\mathbf r_{m}| \approx a_W$, implying $\xi \approx a_W$. 
In the Anderson regime, $\mathbf r_m$ depends highly on the disorder properties, but is usually much less than $a_W$.
In the metallic regime, local maxima of $C(\mathbf r)$ usually appear at $\mathbf r_{m} = 0$, resulting in $\xi = 0$. 

In the main text, we display the phase diagram of attractive disorders with different interaction ranges, as a direct comparison with the STM experiment~\cite{ge2025visualizing}, where we find a qualitative agreement between our results and the experiment. 
There are, however, other parameters that can be changed in our theory, for example, the disorder interaction types (repulsive or attractive). 
Meanwhile, the qualitative agreement with the experiment should be valid in the presence of finite-size effects.
Therefore, it is important to present a detailed analysis of the phase diagram with different system parameters, as we show below.

\begin{figure}
    \centering
    \includegraphics[width=\linewidth]{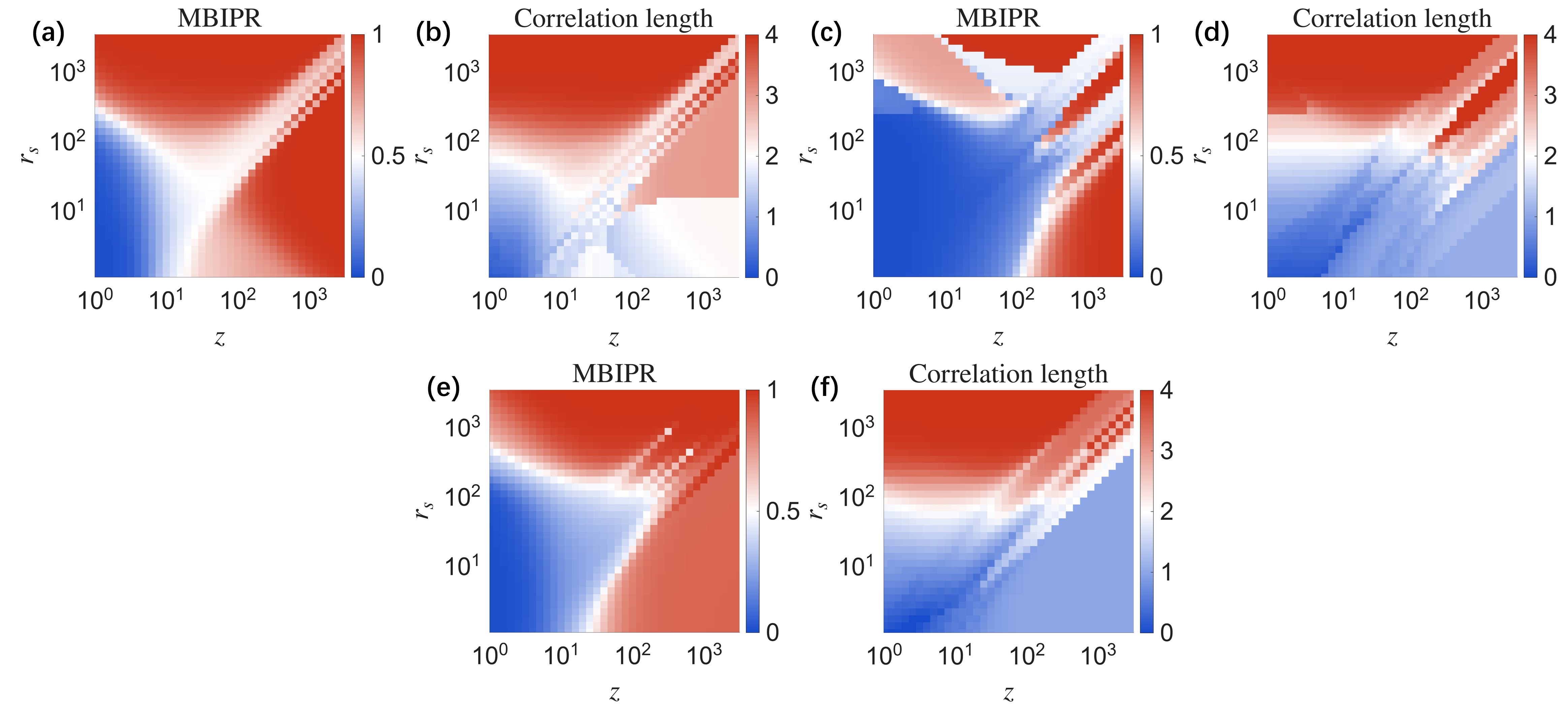}
    \caption{\justifying{Phase diagram for long-range disorders with different interaction types: (a)\&(b) all attractive, (c)\&(d) all repulsive, (e)\&(f) one attractive and one repulsive. For all figures, $N_e = 4, L_x = L_y = 8$, and $N_i = 2$ disorders are located at $(3,4)$ and $(6,6)$.}}
    \label{fig:PD_LR_all}
\end{figure}
\section{\large \uppercase\expandafter{\romannumeral2}.3.1 Phase diagram for different disorder types}
Fig.~\ref{fig:PD_LR_all} shows the phase diagram for different diorder interaction types. 
Similar to the results in the main text, three regimes of the Fermi liquid, Wigner crystal, and Anderson localization are observed, with a sharp transition between the Fermi liquid and the other regimes. And between the WC and Anderson regime, the system experiences a re-entrance where the MBIPR and $\xi$ do not change monotonically, as a result of competition between electron interaction and disorder. 
Notice that the system with repulsive disorders has qualitatively similar behaviors but stronger fluctuations in the ``glass" regime (top right of the phase diagram of Figs.~\ref{fig:PD_LR_all}), which might be due to the different energy landscape induced by different interaction types of disorders. 
% In a clean system, the WC can slide over the system since the background energy is a constant. By adding disorders, the WC is pinned at the minima of the background energy. 
% For attractive disorders, these minima are usually located at the location of disorders with a deep valley. 
% While for repulsive disorders, the energy minima are located away from the disorders' location. 

% Thus, some electrons experience a stronger disorder repulsion since they are closer to the disorders. This may lead to the delocalization of electrons at a smaller $z$, as observed in Fig.~\ref{fig:PD_LR_all}(c).

\begin{figure}
    \centering
    \includegraphics[width=\linewidth]{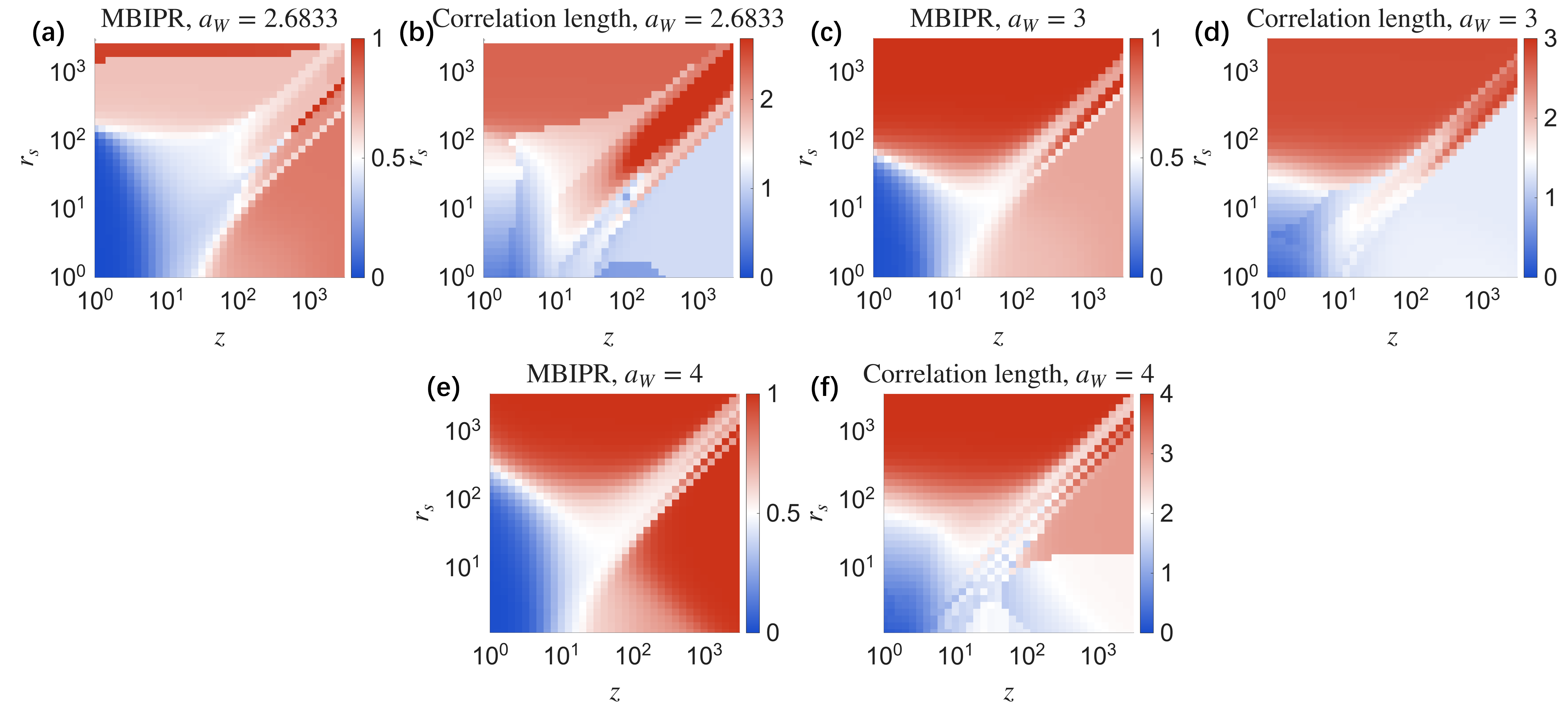}
    \caption{\justifying{Phase diagram for long-range disorders with different Wigner lattice constant $a_W$. For each figure, the electron number and lattice size are respectively: (a)\&(b) $N_e = 5, L_x = L_y = 6$, (c)\&(d) $N_e = 4, L_x = L_y = 6$, (e)\&(f) $N_e = 4, L_x = L_y = 8$. For all figures, $N_i = 2$.}}
    \label{fig:PD_aW_all}
\end{figure}
\section{\large \uppercase\expandafter{\romannumeral2}.3.2 Finite size effects}
Finite-size effects should always exist in our small-sized ED studies. 
To determine whether they are significant in our results, we vary $a_W$ by changing the system size and the number of electrons.
Fig.~\ref{fig:PD_aW_all} shows the phase diagrams for systems with different $a_W$. 
All the phase diagrams show qualitatively similar results, which again confirms our arguments of WC.
Notice the WC regime at $a_W = 2.68$, where the finite-size effect is much stronger, is less smooth than the other two.

\begin{figure}
\centering
\includegraphics[width=0.8\linewidth]{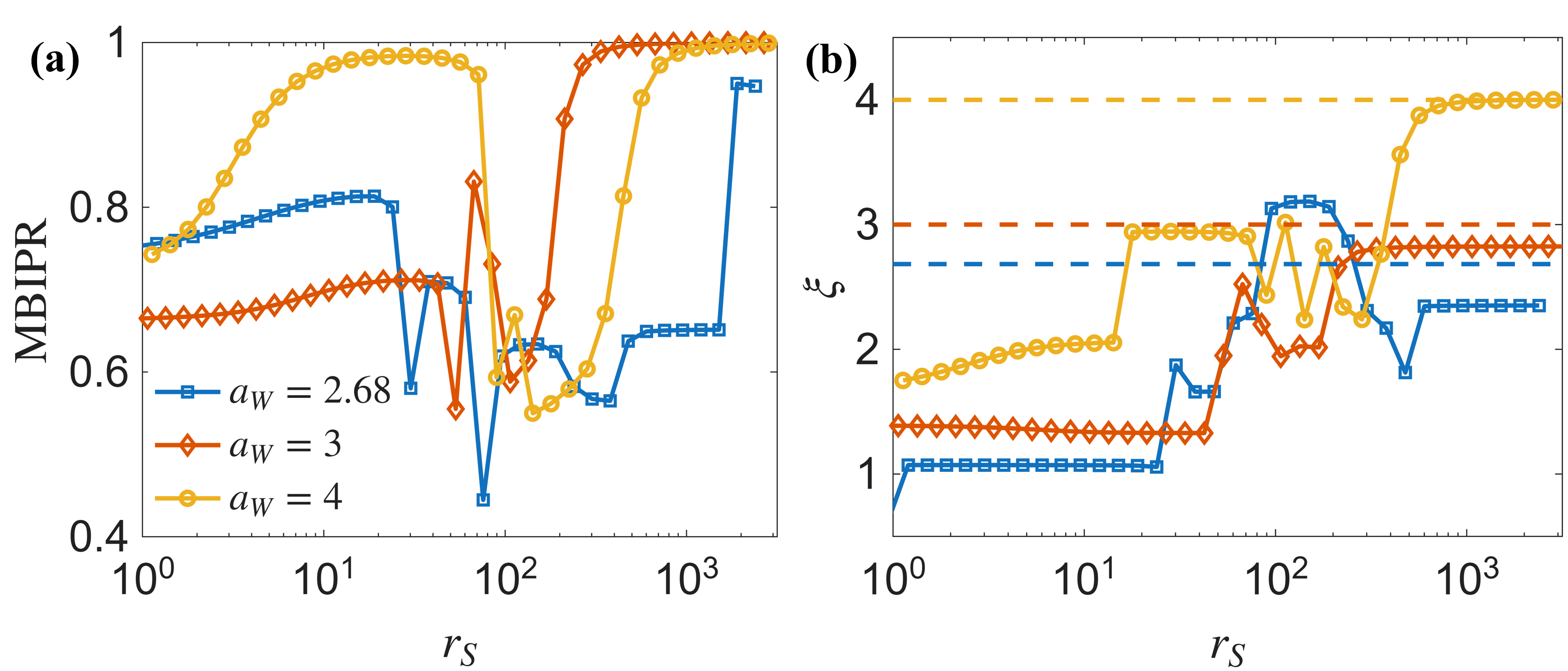}

\captionsetup{justification=raggedright}
\caption{Re-entrant effects at intermediate disorder strength $z\approx 316$ characterized by MBIPR and effective correlation length $\xi$. The dashed lines mark the values of $a_W$. For all figures, $N_i = 2$ .}
\label{fig:cut}
\end{figure}

To take a closer look at the re-entrance effects, we draw a line cut from the phase diagram in Fig.~\ref{fig:PD_aW_all} at $z\approx 316$ (Fig.~\ref{fig:cut}). 
We notice that all systems experience a non-monotonic behavior when increasing $r_s$ from the Anderson to the Wigner regime, analogous to a glassy transition. 
At large $r_s$, $\xi$ approximately saturates to $a_W$. For larger $a_W$, this saturation is almost perfect, which may be due to fewer finite-size effects in a larger system.
Note that more ``plateau" can be observed for $a_W = 2.68$ in both the curves of MBIPR and $\xi$, which could be due to the increase of electrons. 

\end{document}